\documentclass[reprint,amsmath,amssymb,aps,prd]{revtex4-2}

\usepackage{graphicx}
\usepackage{dcolumn}
\usepackage{float}
\usepackage{multirow}
\usepackage{xcolor}
\usepackage{bm}
\usepackage{amssymb}
\usepackage{bm}
\usepackage{hyperref}

\newcommand{\drm}{{\rm d}}

\begin{document}

\title{Adaptive algorithms for low-latency cancellation of seismic Newtonian-noise at the Virgo gravitational-wave detector}

\author{Soumen Koley}
\email{soumen.koley@gssi.it}
\affiliation{Gran Sasso Science Institute (GSSI), I-67100 L'Aquila, Italy}
\affiliation{INFN, Laboratori Nazionali del Gran Sasso, I-67100 Assergi, Italy}
\author{Jan Harms}
\affiliation{Gran Sasso Science Institute (GSSI), I-67100 L'Aquila, Italy}
\affiliation{INFN, Laboratori Nazionali del Gran Sasso, I-67100 Assergi, Italy}
\author{Annalisa Allocca}
\affiliation{Universit\`a di Napoli ``Federico II", I-80126 Napoli, Italy}
\affiliation{INFN, Sezione di Napoli, I-80126 Napoli, Italy}
\author{Francesca Badaracco}
\affiliation{INFN, Sezione di Genova, via Dodecaneso, I-16146 Genova, Italy}
\author{Alessandro Bertolini}
\affiliation{Nikhef, 1098 XG Amsterdam, The Netherlands}
\author{Tomasz Bulik}
\affiliation{Astronomical Observatory, University of Warsaw, Al. Ujazdowskie 4, 00-478 Warsaw, Poland}
\affiliation{Nicolaus Copernicus Astronomical Center, Polish Academy of Sciences, ul. Bartycka 18, 00-716 Warsaw, Poland}
\author{Enrico Calloni}
\affiliation{Universit\`a di Napoli ``Federico II", I-80126 Napoli, Italy}
\affiliation{INFN, Sezione di Napoli, I-80126 Napoli, Italy}
\author{Marek Cieslar}
\affiliation{Nicolaus Copernicus Astronomical Center, Polish Academy of Sciences, ul. Bartycka 18, 00-716 Warsaw, Poland}
\author{Rosario De Rosa}
\affiliation{Universit\`a di Napoli ``Federico II", I-80126 Napoli, Italy}
\affiliation{INFN, Sezione di Napoli, I-80126 Napoli, Italy}
\author{Luciano Errico}
\affiliation{Universit\`a di Napoli ``Federico II", I-80126 Napoli, Italy}
\affiliation{INFN, Sezione di Napoli, I-80126 Napoli, Italy}
\author{Marina Esposito}
\affiliation{Universit\`a di Napoli ``Federico II", I-80126 Napoli, Italy}
\affiliation{INFN, Sezione di Napoli, I-80126 Napoli, Italy}
\author{Irene Fiori}
\affiliation{European Gravitational Observatory (EGO), I-56021 Cascina, Pisa, Italy}
\author{Stefan Hild}
\affiliation{Maastricht University, 6200 MD Maastricht, The Netherlands}
\affiliation{Nikhef, 1098 XG Amsterdam, The Netherlands}
\author{Bartosz Idzkowski}
\affiliation{Astronomical Observatory, University of Warsaw, Al. Ujazdowskie 4, 00-478 Warsaw, Poland}
\author{Alain Masserot}
\affiliation{Universit\'e Savoie Mont Blanc, CNRS, Laboratoire d’Annecy de Physique des Particules - IN2P3, F-74000 Annecy, France}
\author{Beno\^it Mours}
\affiliation{Universit\'e de Strasbourg, CNRS, IPHC UMR 7178, F-67000 Strasbourg, France}
\author{Federico Paoletti}
\affiliation{INFN, Sezione di Pisa, I-56127 Pisa, Italy}
\author{Andrea Paoli}
\affiliation{European Gravitational Observatory (EGO), I-56021 Cascina, Pisa, Italy}
\author{Mateusz Pietrzak}
\affiliation{Nicolaus Copernicus Astronomical Center, Polish Academy of Sciences, ul. Bartycka 18, 00-716 Warsaw, Poland}
\author{Luca Rei}
\affiliation{INFN, Sezione di Genova, via
Dodecaneso, I-16146 Genova, Italy}
\author{Lo\"ic Rolland}
\affiliation{Universit\'e Savoie Mont Blanc, CNRS, Laboratoire d’Annecy de Physique des Particules - IN2P3, F-74000 Annecy, France}
\author{Ayatri Singha}
\affiliation{Maastricht University, 6200 MD Maastricht, The Netherlands}
\affiliation{Nikhef, 1098 XG Amsterdam, The Netherlands}
\author{Mariusz Suchenek}
\affiliation{Nicolaus Copernicus Astronomical Center, Polish Academy of Sciences, ul. Bartycka 18, 00-716 Warsaw, Poland}
\author{Maciej Suchinski}
\affiliation{Astronomical Observatory, University of Warsaw, Al. Ujazdowskie 4, 00-478 Warsaw, Poland}
\author{Maria Concetta Tringali}
\affiliation{European Gravitational Observatory (EGO), I-56021 Cascina, Pisa, Italy}
\author{Paolo Ruggi}
\affiliation{European Gravitational Observatory (EGO), I-56021 Cascina, Pisa, Italy}
\date{\today}
\begin{abstract}
A system was recently implemented in the Virgo detector to cancel noise in its data produced by seismic waves directly coupling with the suspended test masses through gravitational interaction. The data from seismometers are being filtered to produce a coherent estimate of the associated gravitational noise also known as Newtonian noise. The first implementation of the system uses a time-invariant (static) Wiener filter, which is the optimal filter for Newtonian-noise cancellation assuming that the noise is stationary. However, time variations in the form of transients and slow changes in correlations between sensors are possible and while time-variant filters are expected to cope with these variations better than a static Wiener filter, the question is what the limitations are of time-variant noise cancellation. In this study, we present a framework to study the performance limitations of time-variant noise cancellation filters and carry out a proof-of-concept with adaptive filters on seismic data at the Virgo site. We demonstrate that the adaptive filters, at least those with superior architecture, indeed significantly outperform the static Wiener filter with the residual noise remaining above the statistical error bound.
\end{abstract}

\maketitle

\section{Introduction}
\begin{figure}[ht!]
\begin{center}
    \includegraphics[width=0.48\textwidth]{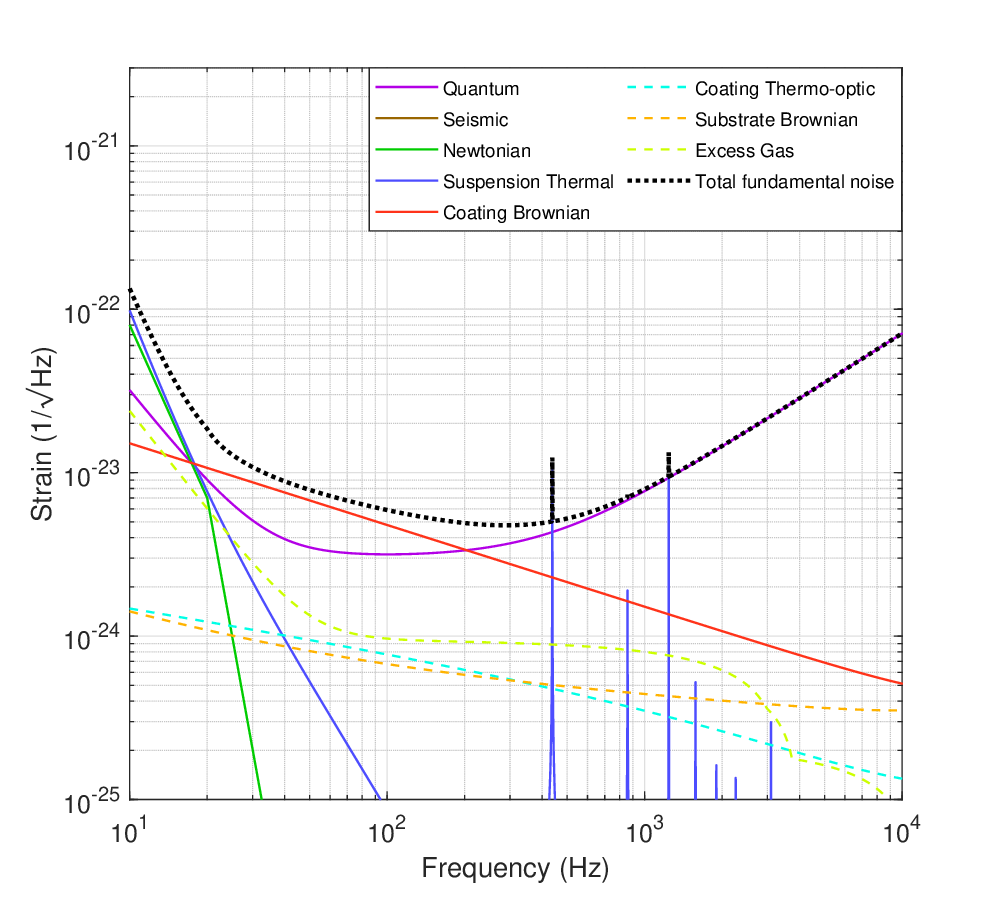}
    \caption{Contribution of several fundamental sources of noise to the AdV+ design sensitivity corresponding to a laser input power of 40\,W and 12\,dB of frequency-dependent squeezing. Newtonian noise is expected to be one of the major contributors to the low-frequency sensitivity}
    \label{AdVPlusDesign}
\end{center}
\end{figure}

Since the first detection of gravitational waves in 2015 \cite{abbott2016observation}, the Advanced Virgo and LIGO detectors \citep{acernese2015advanced,aasi2015advanced} have collectively detected about 90 gravitational wave (GW) signals across three distinct observing runs \citep{abbott2019gwtc, abbott2021gwtc, abbott2023population, abbott2023gwtc}. Between each observing run, phases of technological upgrades are interleaved, targeting enhancements in detector sensitivity and duty-cycle \citep{acernese2019advanced}. The detector's ultimate sensitivity depends on the intrinsic physics embedded within its design, such as laser shot noise at high frequencies \citep{sequino2021quantum} and suspension-thermal noise at low frequencies \citep{hammond2012reducing}. Figure \ref{AdVPlusDesign} shows the contribution of the different fundamental sources of noise to the Advanced Virgo Plus (AdV+) sensitivity. Alongside suspension-thermal noise, Newtonian noise (NN) is anticipated to be a significant obstacle in achieving the desired design sensitivity for frequencies below 20\,Hz. Consequently, one of the planned upgrades for AdV+ before the fourth observing run (O4) involved the design and implementation of a low-latency system aimed at canceling NN \citep{flaminio2020status}. The requirement for low latency comes from certain online analyses providing preliminary parameter estimates, e.g., for masses and sky location.

Newtonian noise arises from the gravitational coupling of terrestrial density fluctuations to the suspended test-masses of the detector \citep{harms2019terrestrial}, which can originate from seismic waves propagating in the subsurface \citep{beccaria1998relevance,hughes1998nn} or variations in pressure and temperature within the atmosphere \citep{creighton2008atm,fiorucci2018impact,brundu2022atm}. In this article, our focus centers on cancellation strategies specifically tailored for seismic NN. Atmospheric NN produced by acoustic noise in the Virgo buildings is predicted to be significantly lower than the targeted AdV+ design sensitivity due to noise-mitigation measures connected to Virgo's air-handling system.

A technique to mitigate NN is through coherent noise cancellation. The NN cancellation (NNC) design and implementation phase follows three key steps: estimating NN to obtain NNC requirements, designing an optimal seismic array layout for cancellation, and implementing algorithms to enable low-latency noise cancellation. Estimation of NN relies on analytical or numerical methods for computing the seismic displacement of the subsurface \citep{harms2019terrestrial,BaderThesis,andric2020sim,harms2022limit}. Simulations necessitate a priori information concerning the seismic properties of the site. This includes the spatial distribution of noise sources near the test-masses, characteristics of the seismic wavefield (whether surface or body waves), and the finite element models representing the infrastructure surrounding the test-masses. Studies utilizing seismic arrays for the decomposition of the wavefield into plane waves have been conducted at Virgo, both inside and outside the End Buildings \citep{koley2017s,tringali2019seismic}. These surveys yield the frequency-dependent direction and velocity of seismic noise propagation at the site. Building on these studies, simulation results for NN estimates concerning Virgo have been detailed in \citep{singha2020fem,singha2021characterization}. For simplicity, the green curve in Figure \ref{AdVPlusDesign} shows the mode of the NN estimate for AdV+. In reality, the NN estimate is dependent on the magnitude of seismic noise and is time-varying.

The next phase involves designing the optimal seismic array layout for NN cancellation. The concept of deploying seismometers around the test-masses to mitigate NN was introduced in \citep{cella2000off}. This approach makes use of the Wiener-Hopf formulation \citep{treitel1970principles}, which establishes a connection between observed seismic displacements and the measured GW strain. Design of such a noise cancellation system have been demonstrated for Advanced LIGO in \citep{driggers2012subtraction, coughlin2016towards,coughlin2018impl}. Adopting a similar strategy, an optimal seismic array designed for NN cancellation was developed for Virgo \citep{badaracco2020machine} and subsequently installed at the Virgo Central Building (CEB), as well as the North and West End Buildings (NEB, WEB). The method performs a global minimization of the frequency domain Wiener residual corresponding to the various array layouts under consideration.

The final step in NN cancellation and the focus of this article involves implementation of algorithms that make use of data from the optimal seismic array (witness channels) to subtract coherent noise from the GW strain data (target channel). In cases of wide-sense stationary inputs, the Wiener filter is the optimal choice for eliminating the contribution of witness channels from the target channel. Standard implementations typically involve computing this filter using extended data periods (lasting days). Long data stretches are used to ensure that the filter coefficients are sufficiently trained to reproduce the data from the target channels. As has been observed for Virgo, static filter of this kind provides sub-optimal cancellation capabilities when handling time-varying inputs \citep{koley2024design}. Addressing this challenge involves recalculating the Wiener filter at designated time intervals. However, implementing this solution in low-latency applications proves impractical due to computational complexity and the ambiguity surrounding the selection of an appropriate time interval for recomputing the filter coefficients. A straightforward approach to tackle this issue is by exploring algorithms that address the Wiener problem and continuously adjust the filter coefficients for incoming samples from the witness channels. Adaptive filtering has been widely used in acoustic echo cancellation, channel equalization, speech processing, and problems related to system identification \citep{haykin2002adaptive}. Inspired by these applications, in this article we explore two classes of algorithms that adaptively solve the minimum-mean-square-error problem. The first one is the least mean square (LMS), which employs a stochastic gradient technique to minimize the mean square of the error signal (Chapter 6 in \citep{benesty2008springer}). The LMS class of algorithms is popular for its computational simplicity. However, its drawback lies in slow convergence and its heavy reliance on the spectral characteristics of witness signals \citep{slock1993convergence}. The second class of algorithms are the recursive least squares (RLS), which solves the quadratic minimization problem exactly at each time step. As demonstrated in \citep{eleftheriou1986tracking}, RLS algorithms exhibit superior tracking behavior over LMS in medium to high signal-to-noise ratio (SNR) environments and are independent of spectral characteristics when it comes to convergence rates. In this article we evaluate the performance and suitability of these two classes of algorithms for implementation as a low-latency NNC at the Virgo GW detector.

The rest of the article is organized as follows: Section \ref{sec:back} presents a background to the cancellation problem and a brief discussion on the seismic environment at Virgo. Fundamental performance limitations of Wiener filters are discussed in Section \ref{sec:limit}. Section \ref{sec:StaticDynamic} compares the noise cancellation performance between static and time-variant Wiener filters, and sets the stage for adaptive filters. Sections \ref{sec:LMS} and \ref{sec:RLS} present the adaptive schemes for the LMS and the RLS filters, and make a quantitative assessment of the subtraction performance of each of the algorithms. Section \ref{sec:FD} addresses limitations in the current methods and explores areas of improvement. Finally the conclusions of the work are presented in Section \ref{sec:conclusion}.

\section{Background}
\label{sec:back}
\begin{figure*}[ht!]
\begin{center}
    \includegraphics[trim={0cm 0.65cm 0cm 0cm},clip,width=\textwidth]{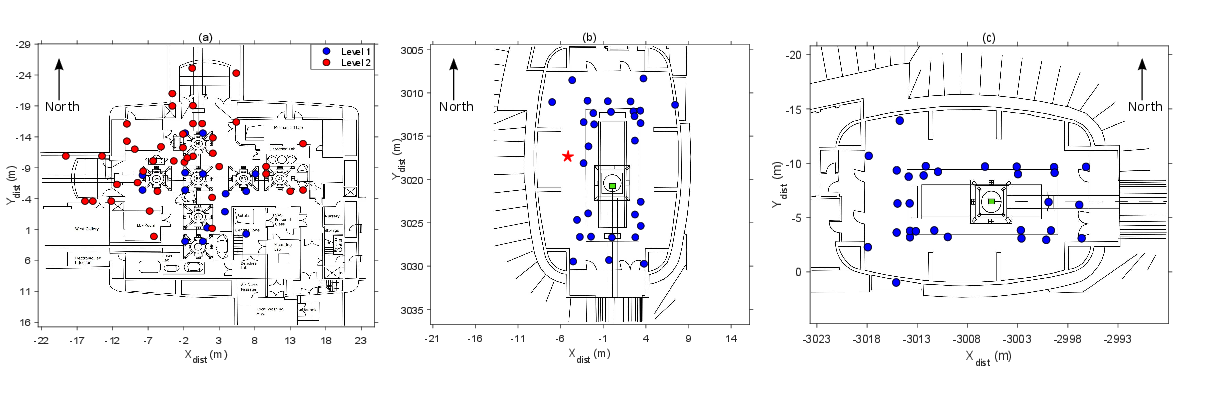}
    \caption{(a) The blue and red solid circles show the positions of the geophones at level 1 and 2 of the CEB, respectively. (b) The blue solid circles show the locations of the geophones and the red star shows the location of the tiltmeter in the NEB. (c) The blue solid circles show the locations of the geophones in the WEB. Note that the origin of the coordinate system corresponds to the location of the beamsplitter at the CEB, and `north' corresponds to the direction of the north arm of the interferometer which is oriented $20^{\circ}$ clockwise with respect to the geographic north}
    \label{NNCArray}
\end{center}
\end{figure*}

\begin{figure}[ht!]
\begin{center}
    \includegraphics[width=0.48\textwidth]{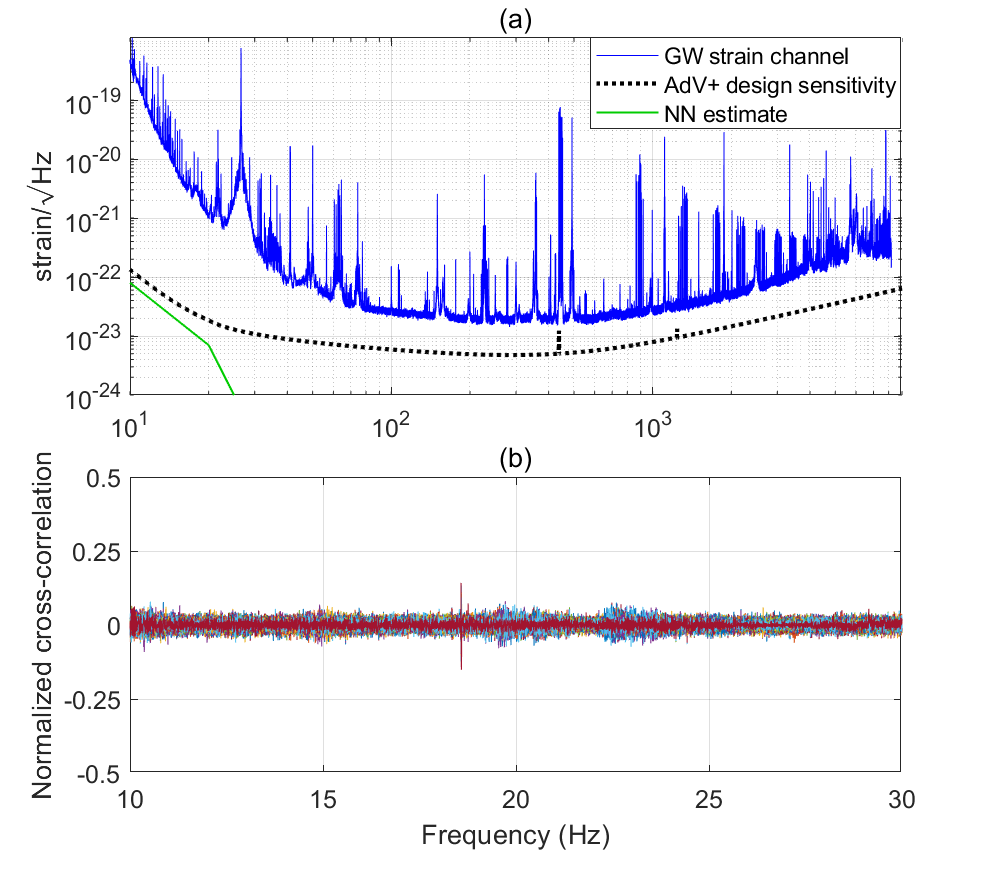}
    \caption{(a) Comparison between AdV+ sensitivity during its commissioning phase before O4 and the design sensitivity. The green curve represents the estimated NN, approximately three orders of magnitude lower than the current sensitivity. (b) Normalized cross-correlations between the 115 geophones and the GW strain channel corresponding to a month of data. No significant cross-correlations are observed.}
    \label{VirgoSens}
\end{center}
\end{figure}

The seismic NN cancellation array in the Virgo GW detector comprises 115 geophones distributed across the CEB (55), NEB (30), and WEB (30). Their locations were determined based on array optimization studies in \citep{badaracco2020machine}. Figures \ref{NNCArray}(a), (b), and (c) depict the seismic arrays at each Virgo building. These geophones have a resonance frequency of 5 Hz and continuously acquire vertical component of the seismic noise at a sampling rate of 500 samples per second. The ideal target channel to showcase the adaptive cancellation of NN is the GW strain output of the interferometer. However, Virgo is presently in its commissioning phase and has yet to achieve its intended sensitivity, particularly within the 10 –- 30\,Hz frequency band, where NN is anticipated to be a major contributor to Virgo's overall noise budget. In Figure \ref{VirgoSens}(a), the AdV+ sensitivity is depicted, derived by averaging the power spectral densities over 300\,s long windows across a full day of data on September 19, 2023. The selection criteria encompassed all time windows where the binary neutron star range exceeded 25\,Mpc. As evident from the figure, Virgo's sensitivity exceeds the design sensitivity by approximately three orders of magnitude. This is further supported in Figure \ref{VirgoSens}(b), which shows that no significant cross-correlations are detected between the NN witness channels and the GW strain channel. The cross-correlations were estimated using the same window lengths and selection criteria of 25\,Mpc as used for generating the power spectral densities.

The seismic wavefield inside the Virgo buildings is mostly dominated by Rayleigh waves \citep{singha2021characterization}. These include both sharp-spectral and broadband sources of noise. Vacuum pumps and motors operating within the Virgo buildings are some of the examples of sharp-spectral noise \citep{TringaliHVAC}. Broadband sources typically originate farther away from the buildings and can be attributed to traffic and farming activities \citep{koley2017s}. In a scenario where the seismic wavefield is primarily dominated by Rayleigh waves, the ground tilt along the direction of the detector is fully coherent with NN from plane Rayleigh waves \citep{harms2016newtonian}. Therefore, a logical alternative to the GW strain channel as the target channel is the tilt signal measured by a tiltmeter. The NEB at Virgo hosts a tiltmeter - which is a highly sensitive prototype balance and was originally developed for the Archimedes experiment \citep{calloni2014towards}. The device is equipped with an interferometric readout and has a resonance frequency of about 23\,mHz which makes it suitable for tilt measurements in the NN band \citep{allocca2021picoradiant}.

\begin{figure}[ht!]
\begin{center}
    \includegraphics[width=0.48\textwidth]{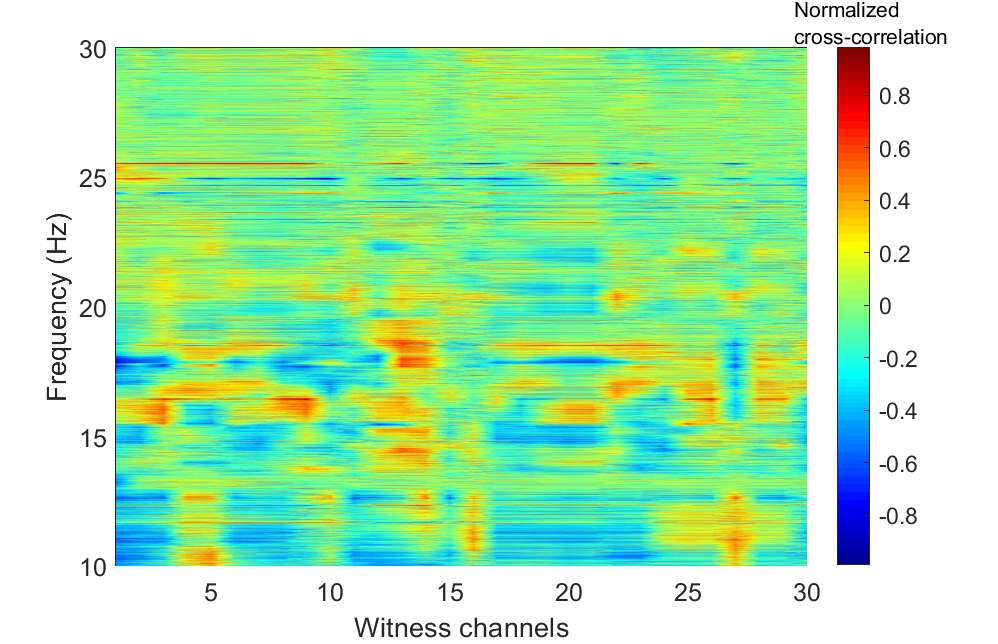}
    \caption{Surface plot showing the cross-correlation between the witness channels and the tilt signal. The colorbar represents the magnitude of the normalized cross-correlations.}
    \label{NEBTiltCC}
\end{center}
\end{figure}

The performance of the noise cancellation system when using the tilt as the target signal strongly depends on the reconstruction accuracy of the tilt signal with the witness channels. An analysis of tilt reconstruction by using the spatial derivative of the vertical component of seismic noise, as measured by an array of geophones, was conducted for Virgo in 2020. Further details can be found in \citep{singha2021characterization}. Another metric to demonstrate cancellation performance using Wiener filters involves estimating cross-correlations between the witness channels and the tilt signal. Figure \ref{NEBTiltCC} shows a surface plot of the normalized cross-correlations between the 30 witness channels and the tilt signal. Strong correlations of about $\pm 0.8$ are observed at frequencies of 11.7, 12.4, 18.5, 23.4, 24.4, and 24.7\,Hz. These correspond to noise sources originating from the heating and the ventilation system at the NEB. Additionally, broadband noise exhibits absolute cross-correlation magnitudes ranging between 0.2 and 0.4. Consequently, strong cancellation is anticipated for the several noise peaks, while the opposite is expected for broadband noise.

\section{Limits of coherent noise cancellation}
\label{sec:limit}
In this section, we calculate two important limits of coherent noise cancellation. We start by considering the static Wiener filter and then discuss the case of adaptive Wiener filters. Generally, the performance of a Wiener filter depends on:
\begin{enumerate}
    \item Number and type of sensors (accelerometer, strainmeter, tiltmeter), which determine the completeness of information we have about the seismic field;
    \item Sensitivity of sensors, which determines the precision of the derived NN model;
    \item Amount of data/information to calculate the Wiener filter, which determines a possible bias of the model.
\end{enumerate}
The Wiener filter can be represented in frequency domain or in time domain. Its coefficients can be assembled in a vector $\mathbf{h}_{n}$, where $n$ is a time index (also applicable in frequency domain where the filter is applied to finite-length data segments), which means that the filter can be time-variant. The filter coefficients are calculated according to
\begin{equation}
\mathbf{h}_{n}= \langle y_n\mathbf{X}_{n}^{\,\dagger}\rangle\cdot \langle \mathbf{X}_{n}\mathbf{X}_{n}^{\,\dagger}\rangle^{-1},
\label{eqWHopf}
\end{equation}
where $y_n$ is a time-domain or frequency-domain sample of the target channel, from which we want to subtract the noise, and $\mathbf{X}_n$ are samples of input channels of the Wiener filter, which provide the coherent information about the noise in the target channel. The vector $\mathbf{X}_n$ can contain more than one sample per channel, e.g., in time domain, it can include multiple past samples of each input channel. The brackets $\langle\cdot\rangle$ indicate an average over many realizations of the noise, and $\dagger$ is the hermitian conjugate, which involves the transposition and a complex conjugate. The term $\langle y_n\mathbf{X}_{n}^{\,\dagger}\rangle$ is the correlation between the target channel and the input channels (a vector of cross-spectral densities in frequency domain) and $\langle \mathbf{X}_{n}\mathbf{X}_{n}^{\,\dagger}\rangle$ is the correlation matrix between all input channels (the cross-spectral density matrix in frequency domain). 

The Wiener filter is the optimal noise-cancellation filter in linear systems. This begs the question whether there are any fundamental limitations to the performance of the Wiener filter. In preparation of such an analysis, let us assume that the estimates $\langle y_n\mathbf{X}_n^{\,\dagger}\rangle,\,\langle \mathbf{X}_n\mathbf{X}_n^{\,\dagger}\rangle$ of the true correlations $\langle y_n\mathbf{X}_n^{\,\dagger}\rangle^0,\,\langle \mathbf{X}_n\mathbf{X}_n^{\,\dagger}\rangle^0$ have errors,
\begin{equation}
\langle y_n\mathbf{X}_n^{\,\dagger}\rangle = \langle y_n\mathbf{X}_n^{\,\dagger}\rangle^0+\bm{\epsilon}_n,\quad
\langle \mathbf{X}_n\mathbf{X}_n^{\,\dagger}\rangle=\langle \mathbf{X}_n\mathbf{X}_n^{\,\dagger}\rangle^0+E_n,
\label{eq:correrr}
\end{equation}
in which case, the corresponding bias of the Wiener filter up to second order in the correlation-estimation errors reads
\begin{equation}
\begin{split}
\mathbf{h}_n&=\mathbf{h}_n^{\,0}+\mathbf{\epsilon}_n\cdot \left(\langle \mathbf{X}_n\mathbf{X}_n^{\,\dagger}\rangle^0\right)^{-1}-\mathbf{h}_n^{\,0}\cdot\left(\langle \mathbf{X}_n\mathbf{X}_n^{\,\dagger}\rangle^0\right)^{-1}\cdot E_n\\
&\;+\mathbf{h}_n^{\,0}\cdot\left(\langle \mathbf{X}_n\mathbf{X}_n^{\,\dagger}\rangle^0\right)^{-1}\cdot E_n\cdot\left(\langle \mathbf{X}_n\mathbf{X}_n^{\,\dagger}\rangle^0\right)^{-1}\cdot E_n\\
&\;-\vec \epsilon_n\cdot\left(\langle \mathbf{X}_n\mathbf{X}_n^{\,\dagger}\rangle^0\right)^{-2}\cdot E_n\\
&\equiv \mathbf{h}_n^{\,0}+\bm{\epsilon}_n^{\;\prime}-\mathbf{h}_n^{\,0}\cdot E_n^{\,\prime}+\mathbf{h}_n^{\,0}\cdot E_n^{\,\prime}\cdot E_n^{\,\prime}-\bm{\epsilon}_n^{\;\prime}\cdot E_n^{\,\prime},
\label{eq:wienererr}
\end{split}
\end{equation}
where in the last line we have subsumed the inverse of the correlation matrix $\langle \mathbf{X}_n\mathbf{X}_n^{\,\dagger}\rangle^0$ into the definition of the bias terms $\bm{\epsilon}_n^{\;\prime},\, E_n^{\,\prime}$. The noise-cancellation procedure is now to multiply the filter coefficients to the samples of the input channels, and the scalar output is the best estimate of the noise contained in the target channel. The procedure leaves the residual $y_n-\mathbf{h}_n\cdot\mathbf{X}_n$. The average power of the residual is given by
\begin{equation}
\begin{split}
&\langle |\mathbf{h}_n\cdot\mathbf{X}_n-y_n|^2\rangle = \langle |\mathbf{h}_n^{\,0}\cdot\mathbf{X}_n-y_n|^2\rangle\\
&\;+\left(\left[\mathbf{h}_n^{\,0}\cdot\langle \mathbf{X}_n\mathbf{X}_n^{\,\dagger}\rangle-\langle y_n\mathbf{X}_n^{\,\dagger}\rangle\right]\cdot \left[\bm{\epsilon}_n^{\;\prime}-\mathbf{h}_n^{\,0}\cdot E_n^\prime\right]^\dagger +{\rm c.c.}\right)\\
&\;+(\bm{\epsilon}_n^{\;\prime}-\mathbf{h}_n^{\,0}\cdot E_n^\prime)\cdot\langle\mathbf{X}_n\mathbf{X}_n^{\,\dagger}\rangle\cdot (\bm{\epsilon}_n^{\;\prime}-\mathbf{h}_n^{\,0}\cdot E_n^\prime)^\dagger\\
&\;+\Big(\left[\mathbf{h}_n^{\,0}\cdot\langle \mathbf{X}_n\mathbf{X}_n^{\,\dagger}\rangle-\langle y_n\mathbf{X}_n^{\,\dagger}\rangle\right]\cdot \left[\mathbf{h}_n^{\,0}\cdot E_n^\prime\cdot E_n^\prime-\bm{\epsilon}_n^{\;\prime}\cdot E_n^\prime\right]^\dagger \\
&\qquad+{\rm c.c.}\Big),
\end{split}
\label{eq:error}
\end{equation}
where $+\rm c.c.$ means to add the complex conjugate of the previous term. The averages $\langle\cdot\rangle$ that appear here are calculated over a different set of noise realizations than the averages that appear in equations (\ref{eq:correrr}) and (\ref{eq:wienererr}). To adopt language from machine learning, one could think of the data used to calculate the Wiener filter in equation (\ref{eq:wienererr}) as training set, and the data used to calculate the average power of the residual in equation (\ref{eq:error}) as the verification set. The result in equation (\ref{eq:error}) has four contributions. The term $\mathbf{h}_n^{\,0}\cdot\langle \mathbf{X}_n\mathbf{X}_n^{\,\dagger}\rangle-\langle y_n\mathbf{X}_n^{\,\dagger}\rangle$, which appears in the second and fourth contribution, is small, i.e., linear in the errors of the correlation estimates calculated from the verification set. This means that the second contribution is really a second-order term in correlation-estimation errors (one error coming from the training set, one from the verification set), and the last contribution is third order. For this reason, we will neglect the contribution to the residual power coming from the last contribution. Since the error in the second contribution depends in an important way on the duration of the verification set, e.g., how many data we use to estimate the power spectral density of residual noise, it is not suited to define a fundamental performance limitation of the Wiener filter. In the following, we will therefore focus on the third contribution, which is quadratic in the errors $\bm{\epsilon}_n^{\;\prime},\,E_n^\prime$ coming from the training set.

\subsection*{Lower bound on noise residuals from statistical errors of the correlation estimates}
Statistical errors of the correlation estimates used to calculate the Wiener filter cause a filter bias. Let us revisit the calculation of the effect of statistical errors on Wiener filtering first presented in \citet{harms2020limit}. A Wiener filter was calculated to find correlations between ground motion and LIGO Hanford GW data. Ground motion was observed with an array of geophones and a tiltmeter. Since the correlated noise in the GW data was expected to be very weak, it was important to assess the statistical significance of the correlation measurement with a Wiener filter. 

Here, we consider the case of a single input channel $X$ to keep the calculation simple, and we assume a frequency-domain Wiener filter and stationary noise so that correlations between frequencies can be neglected, which greatly simplifies the Wiener filter. Then, the statistical estimation error of the cross-spectral density (CSD) $\langle y(f)X(f)^*\rangle$ is 
\begin{equation}
\epsilon(f) = \sqrt{\frac{\langle |X(f)|^2\rangle^0\langle |y(f)|^2\rangle^0}{\nu}}, 
\end{equation}
where $\nu$ is the number of data segments going into the CSD estimate, and $\langle |y(f)|^2\rangle^0$ is the true power spectral density of the target channel $y$. For time-domain filters like the finite-impulse-response Wiener filter, $\nu$ is the product of the duration of the training set used to calculate the filter, and the frequency at which noise is to be subtracted. Inserting this expression into the third contribution on the right-hand-side of equation (\ref{eq:error}), we obtain for the term quadratic in $\epsilon(f)$
\begin{equation}
\boxed{\mathcal B_{\rm stat}(f)=\epsilon^\prime(f)^2\langle|X(f)|^2\rangle \approx \frac{\epsilon(f)^2}{\langle|X(f)|^2\rangle^0}= \frac{\langle|y(f)|^2\rangle^0}{\nu}}
\label{eq:statlimit}
\end{equation}
This bound on subtraction residuals is very powerful since it only depends on the power spectral density of the target channel. It can therefore be easily evaluated for any noise-cancellation scenario.

Similarly, we have for the error $E(f)$,
\begin{equation}
E(f)=\frac{\langle |X(f)|^2\rangle^0}{\sqrt{\nu}},
\end{equation}
and calculating the quadratic term in $E(f)$ of equation (\ref{eq:error}), we find
\begin{equation}
\begin{split}
&|h^0(f)E^\prime(f)|^2\langle|X(f)|^2\rangle \\
&\quad\approx \frac{|h^0(f)E(f)|^2}{\langle|X(f)|^2\rangle^0}= \frac{\langle|h^0(f)X(f)|^2\rangle^0}{\nu},
\end{split}
\end{equation}
where the numerator is the power spectral density of the output of the Wiener filter. This bound therefore requires knowledge of the Wiener filter and is less powerful.

These results can be extended to apply to the case when the input $\mathbf{X}_n$ mapped by the Wiener filter contains $N$ samples per channel and $P$ channels. The statistical error in equation (\ref{eq:statlimit}) increases by a factor $N\cdot P$. In practice, $N$ can be the number of coefficients of a finite impulse response (FIR) filter, which is a few hundred coefficients per channel for Newtonian-noise cancellation, and there can be about 100 seismometers \cite{koley2024design}. This means that the number of averages $\nu$ in equation (\ref{eq:statlimit}) needs to be larger than $10^6$ to be able to achieve noise reduction by a factor 10 in amplitude, which means 10$^5$\,s of data are required in the training set for noise cancellation at 10\,Hz. Let us imagine that someone wants to use such a filter to reduce Newtonian noise at 0.1\,Hz by a factor 100 in amplitude, then $\nu$ needs to be $10^8$ and the training set must have a length of at least $10^9$\,s, which is 30 years. While it is conceivable to have such long training sets in some applications, we will see below that nonstationarities of the seismic field set limits on the Wiener filter performance. One needs to optimize a trade-off between a reduction of the filter bias due to statistical errors as described in this section by increasing the length of the training set, or better tracking the nonstationarities of the field by regularly or continuously updating the filter, which means to reduce the length of the training set. 

\subsection*{Lower bound on noise residuals due to sensor noise}
We can use equation (\ref{eq:error}) to calculate the lower bound on noise residuals from sensor noise. For simplicity, we assume again the case of cancellation of stationary noise with a frequency-domain Wiener filter, but this time using data from $P$ sensors with identical sensor noise with power spectral density $S(f)$. The power spectral density of sensor noise is contained in the diagonal of $\langle \mathbf{X}(f)\mathbf{X}^{\,\dagger}(f)\rangle$. We can consider the contribution of the sensor noise as an error $E(f)$ of the correlation matrix $\langle \mathbf{X}(f)\mathbf{X}^{\,\dagger}(f)\rangle$,
\begin{equation}
E(f)=S(f)\mathbb{I},
\end{equation}
where $\bm{\mathbb{I}}$ is a unit matrix of the same size as the matrix $\langle \mathbf{X}(f)\mathbf{X}^{\,\dagger}(f)\rangle$. Inserting this error term into equation (\ref{eq:correrr}) and equation (\ref{eq:error}), we get
\begin{equation}
\begin{split}
&\langle |\mathbf{h}(f)\cdot\mathbf{X}(f)-y(f)|^2\rangle \\
&= \langle |\mathbf{h}^{0}(f)\cdot\mathbf{X}(f)-y(f)|^2\rangle^0+S(f)\mathbf{h}^{0}(f)\cdot\left(\mathbf{h}^{0}(f)\right)^\dagger\\
&\qquad+S(f)^2\mathbf{h}^{0}(f)\cdot\left(\langle \mathbf{X}(f)\mathbf{X}^{\dagger}(f)\rangle^0\right)^{-1}\cdot\left(\mathbf{h}^{0}(f)\right)^\dagger,\\
&= \langle |y(f)|^2\rangle^0-\mathbf{h}^{0}(f)\cdot\langle\mathbf{X}(f)\mathbf{X}^{\dagger}(f)\rangle^0\cdot\left(\mathbf{h}^{0}(f)\right)^\dagger\\
&\qquad+S(f)\mathbf{h}^{0}(f)\cdot\left(\mathbf{h}^{0}(f)\right)^\dagger\\
&\qquad+S(f)^2\mathbf{h}^{0}(f)\cdot\left(\langle \mathbf{X}(f)\mathbf{X}^{\dagger}(f)\rangle^0\right)^{-1}\cdot\left(\mathbf{h}^{0}(f)\right)^\dagger
\end{split}
\end{equation}
where superscripts $^0$ mean that the sensor noise $S(f)$ is not included in these terms. The result is what we would expect, i.e., the Wiener filter $\mathbf{h}(f)$ maps the sensor noise into the output together with the signal, which leads to an inevitable contribution to the noise residuals:
\begin{equation}
\boxed{\mathcal B_{\rm sens}(f)=S(f)\mathbf{h}^{0}(f)\cdot\left(\mathbf{h}^{0}(f)\right)^\dagger}
\label{eq:snrbound}
\end{equation}
As a simple example, let's assume that all $P$ sensors whose data go into $\mathbf{X}(f)$ see the same signal. In this case, we have
\begin{equation}
\mathbf{h}^{\,0}(f)=\frac{a(f)}{P}(1,\ldots,1),
\end{equation}
where $a(f)$ is the common transfer function from each component of $\mathbf{X}(f)$ to $y(f)$. We then obtain $\mathbf{h}^{0}(f)\cdot\left(\mathbf{h}^{0}(f)\right)^\dagger=|a(f)|^2/P$ and
\begin{equation}
\mathcal B_{\rm sens}(f)=S(f)\frac{|a(f)|^2}{P}
\end{equation}
In this example, the lower bound decreases inversely with increasing number of sensors, which is as expected since the sensor noise effectively averages out over the $P$ sensors forming a more sensitive supersensor of a common signal.

\subsection*{Cancellation limits in the case of nonstationary noise}
Next, we analyze the impact of temporal variations in the plant (e.g., changing correlations) on the statistical bound $\mathcal B_{\rm stat}$. We assume that we have a noise-cancellation filter trying to track the changes in the plant. Such a filter could be an adaptive Wiener filter as described in the subsequent sections or even a time-variant neural network described by linear weights. The number of coefficients describing this filter is $L$ (equal to $N\cdot P$ in the case of a Wiener filter). Now let us assume that the typical time scale of plant variations that we intend to track is $\tau$. Then the maximum number of averages to calculate the filter coefficients for noise cancellation at frequency $f$ is $\nu=f\tau$, and we find the following statistical bound on noise residuals
\begin{equation}
\mathcal B_{\rm stat}(f)\geq \frac{\langle|x(f)|^2\rangle^0}{f\tau}L.
\label{eq:limitnonstat}
\end{equation}
For example, if we want to be able to follow hour-scale variations of the seismic field for NNC at 10\,Hz with a filter that has $L=100$ coefficients, then we can reduce NN at 10\,Hz at most by a factor 19 in amplitude. If we want to follow minute-long variations with an $L=100$ filter, then we can reduce NN at 10\,Hz at most by a factor 2.4 in amplitude. At this point, $\mathcal B_{\rm stat}(f)$ could actually limit the NNC performance, and one might achieve better noise reduction by increasing the averaging time to reduce $\mathcal B_{\rm stat}$ at the cost of not being able to adapt to minute-scale variations in correlations.

There is an important connection between Wiener filtering using sensor data as input and matched filtering of transient events, which is the common technique to model and analyze GW signals. It was shown that in the case of a likelihood analysis of a GW signal and after subtraction of the best-fit $\hat y(f)$ from the GW data that includes the true signal $\tilde y(f)$, a residual is left in the data whose SNR (in power) is $L$, where $L$ is the number of parameters of the generally nonlinear signal model \cite{cutler2006background},
\begin{equation}
4\int\limits_0^\infty\drm f\frac{\langle|\tilde y(f)-\hat y(f)|^2\rangle}{S(f)}=L,
\label{eq:restransient}
\end{equation}
where the numerator under the integral contains the average over many transients subtracted with this model. Hence, the number of filter parameters appears equally as factor in the statistical bound of linear Wiener filtering and of nonlinear matched-filter-based transient subtraction. In fact, the expressions in equations (\ref{eq:restransient}) and (\ref{eq:limitnonstat}) are analogous with two important differences. The first difference is that Wiener filters can be improved by averaging over many realizations of the noise leading to a reduction of the bound by a factor $\nu=f\tau$. In matched-filter analyses, a model needs to be matched to an observed transient under variation of parameters with ad hoc unknown values, and this must be done for every transient individually. This is true for noise transients as for GW signals with the additional burden that one must provide a faithful model of a noise transient \cite{payne2022transient}. The second difference is that the model used for the matched-filter transient subtraction can accumulate information about the transient from different frequencies, which means that for transients with broad spectrum, the residual at each frequency might lie below other instrument noise. The only way to further reduce the residual left by a transient is if \emph{exactly the same transient repeats} and therefore the model of the transient can be gradually improved. Also note that if there is uncertainty of the occurrence time of a transient with known shape, then $L=1$, and this is enough to enforce the SNR bound on the residual after subtraction. In practice for NNC, the residuals of such transients would be higher. It should be stressed that other limiting factors like incomplete information about the seismic field or systematic errors can become relevant before the ultimate performance limitations described in this section are reached.

\section{Static and Dynamic Wiener Filter}
\label{sec:StaticDynamic}
The Virgo NN cancellation problem is envisaged as a low-latency Multiple Input Single Output system. At any time index $n$, the past $N$ samples from $P$ witness channels are used to compute an estimate of a target sample $y_n$. The Wiener filter coefficients $\mathbf{h}_n$ are obtained by minimizing the error 
\begin{equation}
    e_n \triangleq y_n - \mathbf{h}_n\mathbf{X}_n,
    \label{eqn1}
\end{equation}
where $\mathbf{X}_n = [\mathbf{x}^{\dagger}_n,\mathbf{x}^{\dagger}_{n-1}, \cdots, \mathbf{x}^{\dagger}_{n-N+1}]^{\dagger}$ is a $(NP\times1)$ column vector of the data from the witness channels, and $\mathbf{x}_n = [x_n^1, x_n^2, \cdots, x_n^P]^{\dagger}$. As stated previously in equation (\ref{eqWHopf}), the optimal time domain filter $\mathbf{h}^0_n$ is obtained by solving $\frac{\partial E\{e^2_n\}}{\partial \hat{\mathbf{h}}_n} = \mathbf{0}$, which leads to the Wiener-Hopf solution
\begin{equation}
    \mathbf{h}^0_n = \mathbf{P}\mathbf{R}^{-1},
    \label{eqn2}
\end{equation}
where for simplicity and future presentation we denote $\langle \mathbf{X}_{n}\mathbf{X}_{n}^{\,\dagger}\rangle$ with $\mathbf{R}$ and $\langle y_{n}\mathbf{X}_{n}^{\,\dagger}\rangle$ with $\mathbf{P}$. In our application, seismic noise measured by the different witness channels is correlated. Consequently, the matrix $\mathbf{R}$ is rank deficient within numerical precision. Hence, we seek a solution as
\begin{equation}
    \min\{\|\mathbf{h}_n\|_2 \mid \mathbf{h}_{n}\mathbf{R} = \mathbf{P}\}
    \label{eqn3}
\end{equation}
where $\|\cdot\|_2$ represents the Euclidean norm.

The NN cancellation system targets noise within the 10 -- 25\,Hz frequency band. Hence, prior to estimating the noise-cancellation filters a series of pre-processing steps are implemented. These include downsampling the witness and target channel data to 100\,Hz, followed by bandpass filtering within the 10 -- 25\,Hz range. All the filters used in pre-processing are FIR filters. This ensures causality of the noise subtraction process. Data from each of the witness and the target channels are then linearly detrended and scaled using the standard deviation specific to each channel's data. Other parameters that need to be determined before estimating and applying the Wiener filter to longer data stretches are the number of witness channels $P$ and the filter length $N$. Amongst the 30 channels at the NEB, we select 24 for our analyses. Given the proximity of several channels within a meter of each other, we excluded six channels from the analysis. Ideally, all channels should be included, but special frequency-dependent processing is required to deal with high correlations between close seismometers, which we avoid in our first analysis of adaptive-filter performances. While correlations are very high between all close sensor pairs below 15\,Hz, architectural features of the Virgo NEB and WEB cause the correlation to fall above 15\,Hz between some of the sensors even if they are only separated by one meter \cite{tringali2019seismic}.

\begin{figure}[ht!]
\begin{center}
    \includegraphics[width=0.48\textwidth]{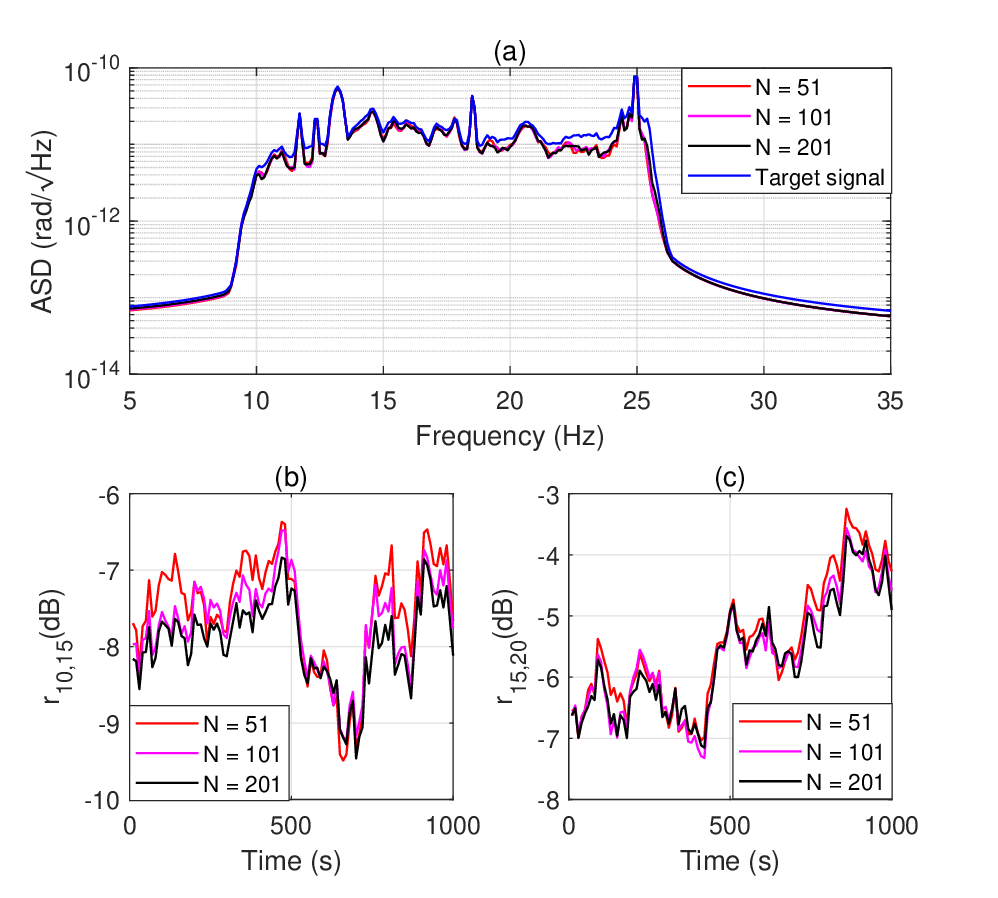}
    \caption{(a) Amplitude spectra of the target and Wiener-reconstructed signals computed by averaging over 10\,s windows within a 1000\,s data stretch. (b) Noise cancellation in the 10 -- 15\,Hz frequency band corresponding to three different Wiener filter lengths. (c) same as (b), but corresponds to the 15 -- 20\,Hz band.}
    \label{FiltLength}
\end{center}
\end{figure}

The choice of the number of filter coefficients $N$ per witness channel is not trivial. Choosing a too low number of filter coefficients introduces a bias in the estimate of the target sample. This bias depends on the energy of the omitted terms of the filter coefficients and the cross-correlations between witness channels. For the latter, an understanding of the physics of the system plays a key role. In our particular case where the witness channels are separated maximally by about 30\,m with seismic-wave propagation speeds between 100 and 150\,m/s \citep{singha2021characterization,koley2024design}, we do not expect significant cross-correlations between the witness channels beyond half a second ($\approx$ 50 samples at a sampling frequency of 100\,Hz). Although, it might be possible that reflected seismic waves or their reverberations are visible in the cross-correlations at a much later time. Hence, we performed a test of the noise cancellation performance corresponding to three different filter lengths of $N=51, 101, 201$. Figure \ref{FiltLength}(a) shows the amplitude spectrum of the target signal alongside the Wiener-reconstructed signals for these three filter lengths. The amplitude spectra of the reconstructed signals exhibit similarity, with subtle differences observed in cancellation performance. We define the cancellation performance in decibels (dB) for the frequency band $f_1 - f_2$ as
\begin{equation}
    r_{f_1,f_2} = 10\times \log_{10}\left(\frac{\sum\limits_{f=f_1}^{f_2} e^2(f)}{\sum\limits_{f=f_1}^{f_2} y^2(f)}\right),
    \label{eqn4}
\end{equation}
where $e(f)$ and $y(f)$ represent the absolute values of the Fourier transforms of the error and the target signals, respectively. The Fourier transforms are evaluated every 10\,s. In Figures \ref{FiltLength}(b) and (c), the cancellation performance over a 1000\,s data duration is presented. As mentioned previously, the cancellation performance remains within a dB of each other and the cancellation performance improves marginally for increasing filter lengths. It is also worth noting that with increasing number of filter coefficients, the complexity of the linear system in equation (\ref{eqn2}) increases and the conditioning of the matrix $\mathbf{R}$ worsens. Hence, adhering to both computational and physical constraints we chose $N=101$ for our analysis.

The next parameters that need to be decided upon are the length of the data stretches needed to calculate the Wiener filter coefficients, and how often the filter coefficients need to be reevaluated to adapt to potential changes in the seismic field. In the case when the noise statistics are stationary, reevaluation of the filter coefficients is not necessary. The choice of these window lengths depends on two factors. The first is the statistical limit as discussed in Section \ref{sec:limit}. This means that the noise cancellation performance of the Wiener filter can potentially be limited by statistical errors if the data stretch is not long enough. The second one is the SNR-bound given in equation (\ref{eq:snrbound}). If the seismometers measure ground displacements with an SNR of 20, the best cancellation performance that can be achieved with respect to the SNR-bound is about 13\,dB (can be better or worse depending on the correlations between seismometers). Therefore in such cases, even if the longer stretches of data are used to calculate the Wiener filters, the cancellation performance cannot be improved beyond the SNR-bound. In order to understand the effect of these parameters on the noise cancellation performance, we define two cases of Wiener-filter evaluation. The first one we refer to as the static Wiener filter (SWF). This corresponds to the case, when a fixed length of data is used to calculate the filter coefficients, and the coefficients do not change with time. The same filter coefficients are used to cancel noise for all subsequent data. The second one we refer to as the updated Wiener filter (UWF). Unlike the SWF, the filter coefficients of the UWF are reevaluated after a given time interval. In the next few paragraphs, we present a comparative analysis of the noise cancellation performance of the SWF implementation and a Wiener filter that is regularly updated.

\begin{figure}[ht!]
\begin{center}
    \includegraphics[width=0.48\textwidth]{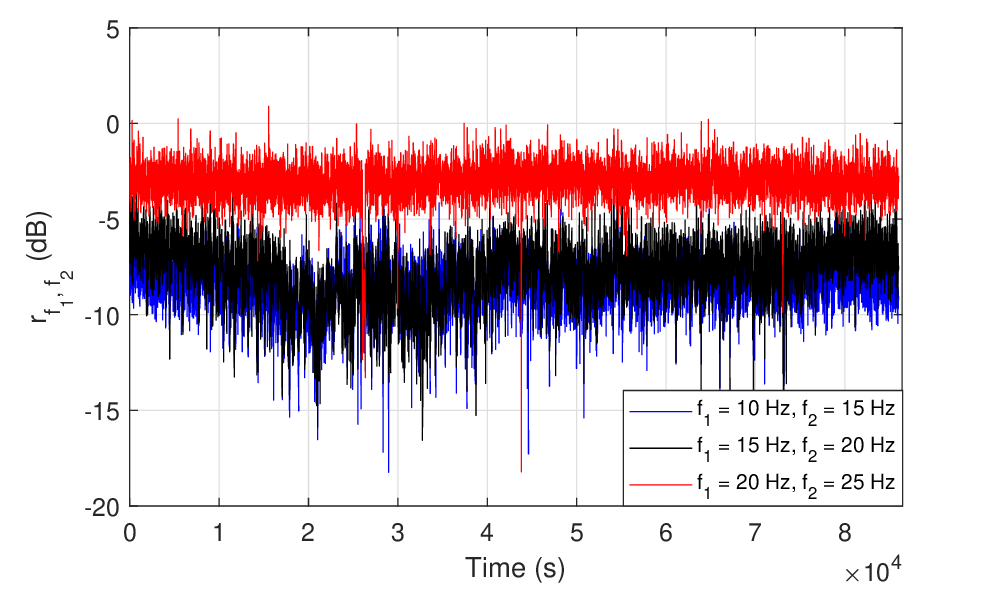}
    \caption{Noise cancellation performance of the UWF reevaluated every 1000\,s corresponding to a full day of data from August 01, 2023. The blue, black, and the red curves show the cancellation performance for the frequency bands: 10 -- 15\,Hz, 15 -- 20\,Hz, 20 --25\,Hz, respectively.}
    \label{UWF1000s}
\end{center}
\end{figure}

\begin{figure}[ht!]
\begin{center}
    \includegraphics[width=0.48\textwidth]{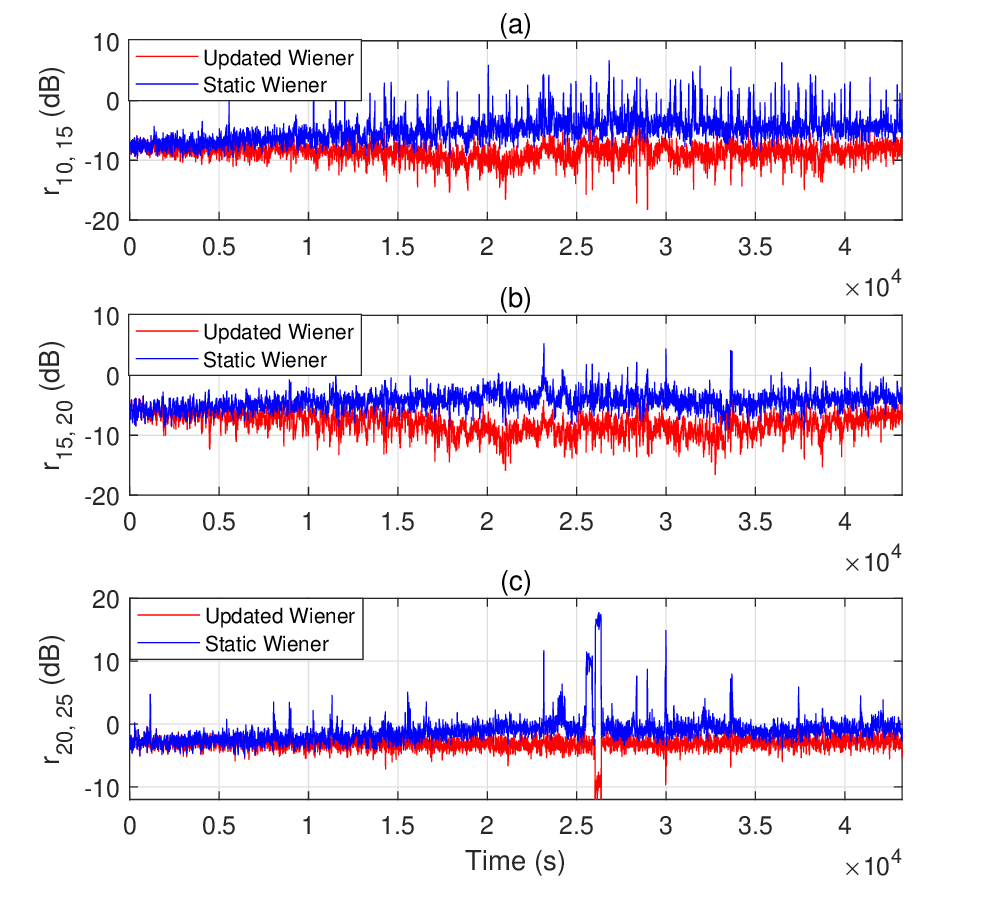}
    \caption{Comparison of noise cancellation performance using the SWF (blue curve) computed from a 1000\,s data starting at 00:00:00 UTC on August 01, 2023, applied to subsequent data stretches for the same day, and the UWF (red curve) estimated every 1000 s on August 01, 2023, across frequency bands: (a) 10 -- 15\,Hz, (b) 15 -- 20\,Hz, and (c) 20 -- 25\,Hz.}
    \label{StaticDynamic1000s}
\end{center}
\end{figure}

We calculate the SWF and UWF for two different scenarios. In the first scenario, we estimate the Wiener filter using data from the witness and target channels for a period of 1000\,s starting at 00:00:00 UTC on August 01, 2023. The estimated SWF was applied to the full subsequent data stretch of the same day. The UWF coefficients were reevaluated every 1000\,s. This last setup operates like an offline cancellation scheme. Filter coefficients are estimated every 1000\,s of data and subsequently applied to the same 1000\,s of data from the witness channels to cancel the target signal. The choice of the 1000\,s interval for evaluating filter coefficients was made to keep $\mathcal B_{\rm stat}$ well below the observed filter performance. Before applying the filters to the data, we used the same detrending and scaling coefficients used during the filter calculation process. Figure \ref{UWF1000s} illustrates the noise cancellation performance of the UWF reevaluated every 1000\,s corresponding to the frequency bands 10 -- 15\,Hz, 15 -- 20\,Hz, and 20 -- 25\,Hz. A cancellation between 10 -- 15\,dB is observed for the frequency band 10 -- 20\,Hz and between 5 -- 10\,dB in the 20 -- 25\,Hz band. Following equation (\ref{eq:statlimit}), the statistical limit to the noise cancellation is between 15 -- 20\,dB. Therefore, making use of more data like 2000\,s to reevaluate the filter coefficients will not improve the cancellation performance, which is instead limited by something else. The limit is probably not coming from the sensor SNR either since ground displacement inside the Virgo buildings in the 10\,Hz -- 25\,Hz band is strong leading to high SNR. It is possible that the seismometers do not provide the full required information about the seismic field to model the noise in the target channel, or there are temporal variations of the seismic field that the filter is not able to track by updating it every 1000\,s. 
Figures \ref{StaticDynamic1000s}(a), (b), and (c) illustrate some of the disadvantages of using the SWF. Both filters show comparable performance up to a few thousand seconds post the SWF estimation. However, a gradual degradation in performance of the SWF is observed as time approaches mid-day. This is due to the fact that the SWF was estimated using data from a quiet time around mid-night on August 01, 2023. Consequently, the static filter performs worse compared to the UWF during the day time. Several instances when $r_{10,15}$ exceeds 0\,dB are also observed. This implies that the SWF occasionally adds noise instead of subtracting noise from the target channel.

In the second scenario, we estimated the SWF by using a day of data from the witness and target channels. This test was done to verify if the cross-correlations estimated using a day of data provide a better average performance over a full day compared to using 1000\,s of data from night time. These filters were then applied to cancel the target signal from the next days. The correlations between the data from the witness channels were first computed every 1000\,s and then averaged across all such windows within a day of data. The averaged cross-correlations were used to create the matrix $\mathbf{R}$. Similarly, the row vector $\mathbf{P}$, which comprises the cross-correlations between the witness and target channels, was populated by estimating cross-correlations every 1000\,s and averaging them over a full day. The steps for estimating the UWF were the same as stated under the first scenario.
 
\begin{figure}[ht!]
\begin{center}
    \includegraphics[width=0.48\textwidth]{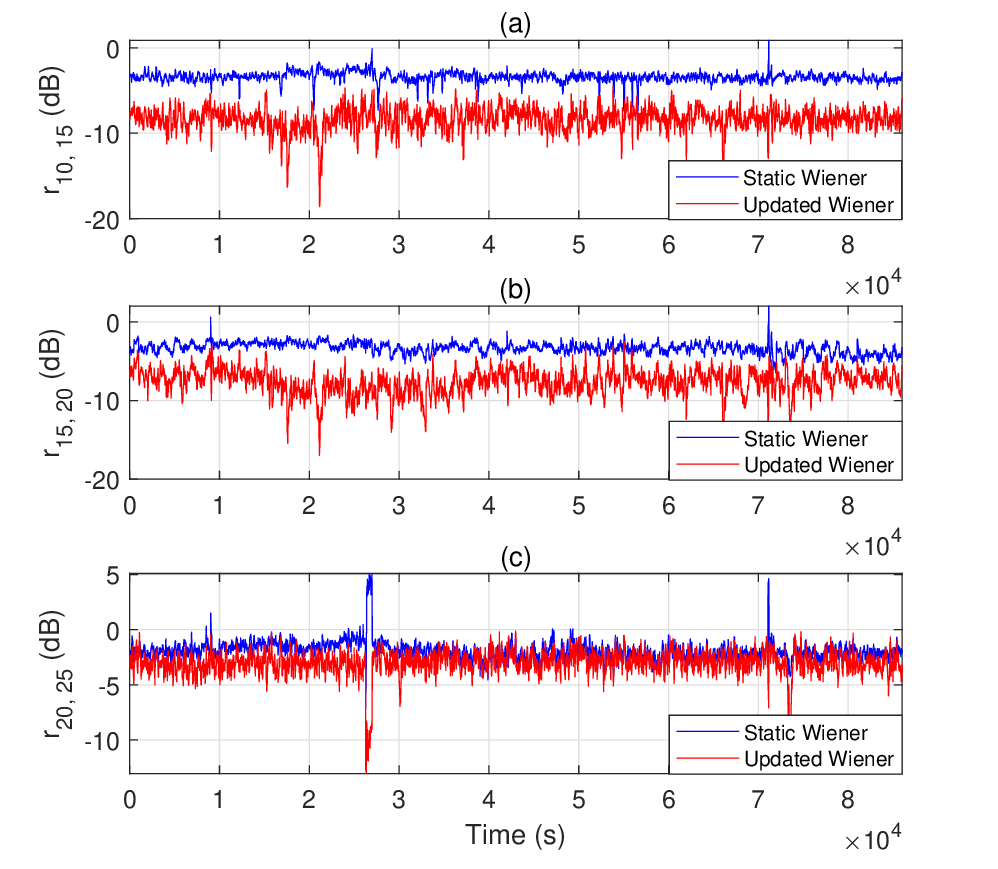}
    \caption{Comparison of noise-cancellation performance using static Wiener filter (blue curve) computed from a full day's data on July 31, 2023, applied to August 01, 2023, and UWF (red curve) updated every 1000\,s on August 01, 2023, across frequency bands: (a) 10 -- 15\,Hz, (b) 15 -- 20\,Hz, and (c) 20 -- 25\,Hz.}
    \label{StaticDynamic}
\end{center}
\end{figure}
Similar to the comparison presented in Figure \ref{StaticDynamic1000s}, the performance comparison for the daily-averaged SWF and UWF are shown in Figures \ref{StaticDynamic}(a), (b), and (c). For this particular implementation, the SWF coefficients were estimated from data on July 31, 2023 and applied to data from the following day (August 01, 2023). Unlike the scenario where the static Wiener filter was estimated using a 1000\,s stretch of data (Figure \ref{StaticDynamic1000s}), the performance of the daily-averaged static Wiener filters does not exhibit diurnal variation. A decrease in the percentage of time when the SWF introduces noise to the target channels is also observed. The stability in performance of the daily-averaged static Wiener filter can be attributed to it being calculated using a day of data, as opposed to just a 1000\,s duration. During the day time when the SNR of the transients increase and a good cancellation is observed with the UWF, the SWF shows approximately 5\,dB lower cancellation performance in the 10 -- 15\,Hz and 15 -- 20\,Hz bands.

Overall, for both scenarios presented above, the performance of the SWF lags behind the UWF. This underscores the necessity for adaptive filters, a topic to be discussed in upcoming sections. In order to assess the noise-cancellation performance of the adaptive filters, the UWF evaluated every 1000\,s will be used as a benchmark. This seems a good choice, given the Wiener formulation of the subtraction scheme.

\section{Least Mean Square algorithm}
\label{sec:LMS}
The Least Mean Square (LMS) algorithm \citep{widrow1960adaptive} derives its roots from the steepest descent algorithm. In the steepest descent algorithm, the filter coefficients at time index $n$ are adapted as
\begin{equation}
    \mathbf{h}_{n} = \mathbf{h}_{n-1} + \mu(\mathbf{P}-\mathbf{h}_{n-1}\mathbf{R}),
    \label{eqn5}
\end{equation}
where $\mu>0$ is the step-size parameter. However, computing the matrix $\mathbf{R}$ and the row vector $\mathbf{P}$ at every new time index is computationally inefficient and not suitable for low-latency applications. Hence, the LMS method which is a stochastic gradient algorithm uses the instantaneous estimates of $\mathbf{R}$ and $\mathbf{P}$. The filter coefficients are then adapted as
\begin{align}
    \mathbf{h}_{n} &= \mathbf{h}_{n-1} + \mu \mathbf{X}^{\dagger}_{n}\left(y_n-\mathbf{h}_{n-1}\mathbf{X}_{n}\right) \nonumber \\
    \implies \mathbf{h}_{n} &= \mathbf{h}_{n-1} + \mu \mathbf{X}^{\dagger}_{n}\mathcal{E}_{n} \label{eqn6}
\end{align}
where $\mathcal{E}_{n} = y_n-\mathbf{h}_{n-1}\mathbf{X}_{n}$. The condition for convergence of $\mathbf{h}_{n}$ to the optimal filter coefficients for a given system is $0<\mu<\frac{2}{\lambda_{\text{max}}}$, where $\lambda_{\text{max}}$ is the maximum eigenvalue of the matrix $\mathbf{R}$ (equation 6.8 in \citep{widrow1985adaptive}). With minor modifications to equation (\ref{eqn6}), the normalized version of the LMS algorithm is derived such that the step-size can be expressed independent of the eigenvalue-spread of matrix $\mathbf{R}$.

\subsection{Normalized Least Mean Square}
\begin{figure}[ht!]
\begin{center}
    \includegraphics[width=0.48\textwidth]{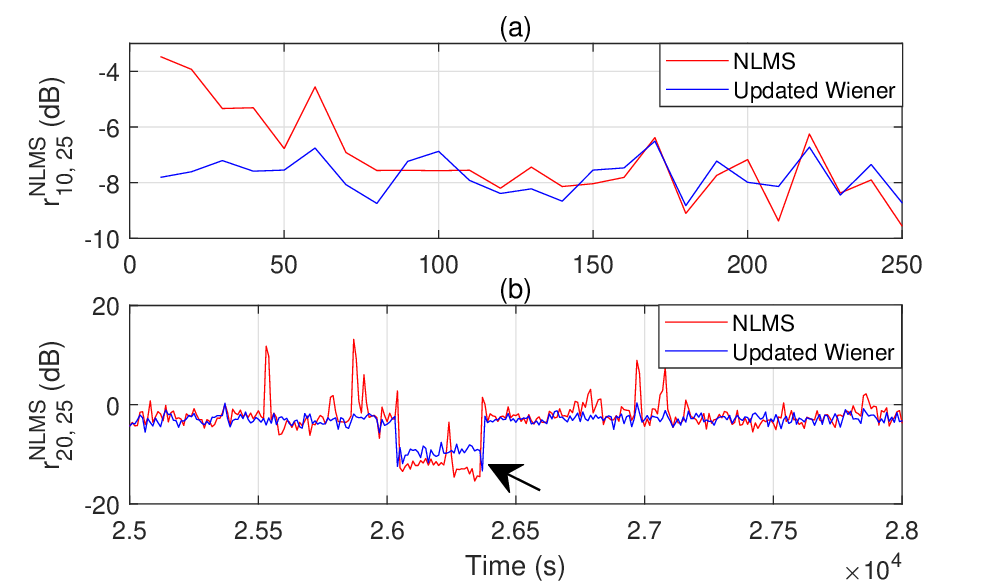}
    \caption{(a) The red curve shows the convergence of the NLMS to the UWF cancellation performance (blue curve) within the first 80 -- 100\,s after start. (b) Comparison of the transient noise cancellation performance of the NLMS to the UWF in the frequency band 20 -- 25\,Hz. The arrow points to the time stretch when the NLMS filter adjusts itself to match the spectral characteristics of the transient.}
    \label{NLMSConvTrack}
\end{center}
\end{figure}

The Normalized LMS (NLMS) adaptation can be expressed as
\begin{equation}
    \mathbf{h}_{n} = \mathbf{h}_{n-1} + \alpha\frac{\mathbf{X}^{\dagger}_{n}\mathcal{E}_{n}}{\mathbf{X}^{\dagger}_{n}\mathbf{X}_{n}+\delta_{\text{NLMS}}}, 
    \label{eqn7}
\end{equation}
where $0<\alpha<2$ is the normalized step-size, and $\delta_{\text{NLMS}}$ is a small positive constant to avoid division by zero in case when the input signals are zero vectors. From equation (\ref{eqn7}), it is evident that the selection of $\alpha$ is crucial for the filter adaptation. The NLMS algorithm demonstrates the fastest convergence when $\alpha = 1$ \citep{makino1993exponentially}. Consequently, we adopted $\alpha = 1$, initializing the algorithm with $\mathbf{h}_{-1} = \mathbf{{0}}_{1\times NP}$. Data pre-processing for the witness and target channels were similar to the ones during the UWF implementation. Figure \ref{NLMSConvTrack}(a) depicts the NLMS algorithm's convergence to the noise cancellation performance of the UWF. Typically, the NLMS algorithm converges at a rate of about 20\,dB per $5NP$ samples for white inputs \citep{rupp1998family}. However, in our correlated input scenario, convergence might take longer. For instance, with $N=101$ and $P = 24$, we observed convergence to approximately 10\,dB in about $5NP$ samples (at a sampling rate of 100\,Hz). In Figure \ref{NLMSConvTrack}(b), a comparison of the transient-tracking performance of the NLMS algorithm is depicted. The arrow in the figure highlights a transient lasting approximately 500\,s. True to the cause, we observe that the NLMS algorithm performs noise cancellation which is comparable to the UWF method.

\begin{figure}[ht!]
\begin{center}
    \includegraphics[width=0.48\textwidth]{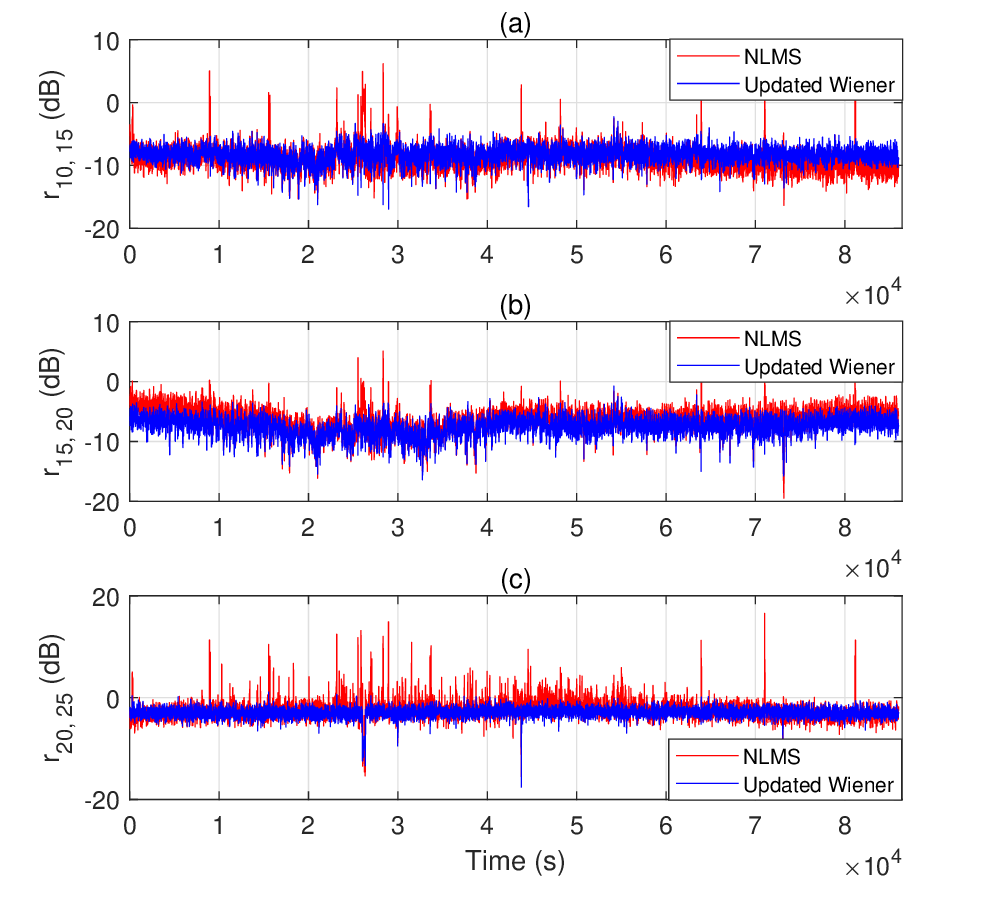}
    \caption{Comparison of noise cancellation performance between the NLMS algorithm (red curve) and the UWF method (blue curve) for August 01, 2023 data, across frequency bands: (a) 10 –- 15\,Hz, (b) 15 -- 20\,Hz, and (c) 20 –- 25\,Hz.}
    \label{NLMSUWFCompare}
\end{center}
\end{figure}

In order to compare the performance of the NLMS algorithm against the UWF over longer time scales, the NLMS algorithm was run on seven days of continuous data (August 01 -- August 07, 2023). Figures \ref{NLMSUWFCompare}(a), (b), and (c) show the noise-cancellation performance comparison between the NLMS and UWF methods for a single day (August 01, 2023). The NLMS method performs as well or slightly better compared to the UWF in the 10 -- 15\,Hz and 20 -- 25\,Hz bands. However, an offset of about 2\,dB is observed in the 15 -- 20\,Hz band. This is a characteristic spectral bias often seen with stochastic gradient methods \citep{eleftheriou1986tracking}. Across the entire seven-day analysis window, we calculated the performance difference between the NLMS and UWF methods within the frequency band $f_1 - f_2$ as
\begin{equation}
    \mathcal{D}^\text{NLMS}_{f_1, f_2} = r^{\text{NLMS}}_{f_1, f_2} - r^{\text{UWF}}_{f_1, f_2}.
    \label{eqn8}
\end{equation}

\begin{figure}[ht!]
\begin{center}
    \includegraphics[width=0.48\textwidth]{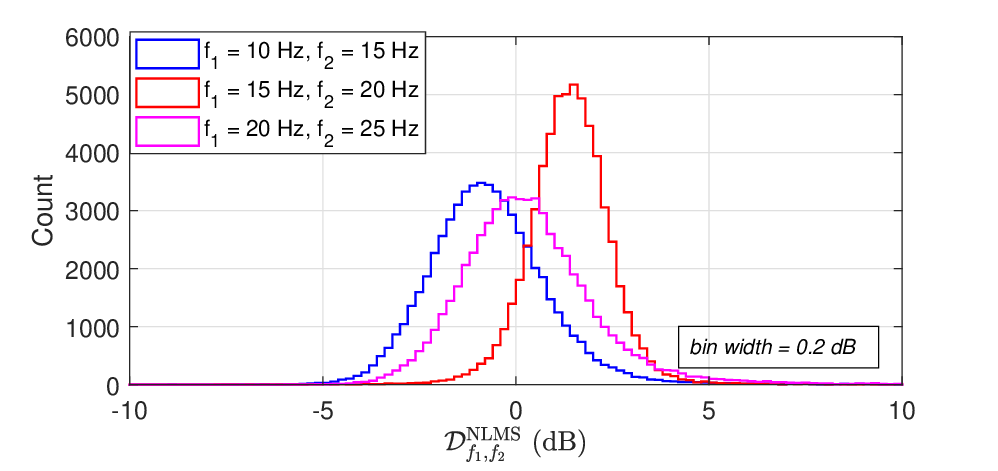}
    \caption{Comparison of the histograms of $\mathcal{D}^{\text{NLMS}}_{f_1, f_2}$ corresponding to the frequency bands 10 -- 15\,Hz, 15 -- 20\,Hz, and 20 -- 25\,Hz.} 
    \label{HistNLMSUWF}
\end{center}
\end{figure}

\noindent
Figure \ref{HistNLMSUWF} displays the histograms of $\mathcal{D}^{\text{NLMS}}_{f_1, f_2}$ for three distinct frequency bands: 10 -- 15\,Hz, 15 -- 20\,Hz, and 20 -- 25\,Hz. Notably, the NLMS outperforms the UWF in the 10 -- 15\,Hz band approximately 70\% of the time. Conversely, in the 15 -- 20\,Hz band, the UWF surpasses the NLMS in performance for about 90\% of the time with a positive mean in the distribution. For the 20 -- 25 \,Hz band, the mean of the distribution is close to zero with an almost equal percentage of data on both sides. An observation to note is that $\mathcal{D}^{\text{NLMS}}_{f_1,f_2}<0$ indicates the NLMS outperforms the UWF, and vice versa for $\mathcal{D}^{\text{NLMS}}_{f_1,f_2}>0$. We also assess the percentage of time when $\mathcal{D}^{\text{NLMS}}_{f_1,f_2}\leq2$. The latter implies that the NLMS performance is at most 2\,dB worse than the UWF. Another crucial metric is the percentage of time when $r^{\text{NLMS}}_{f_1, f_2} >0$. This indicates instances where the NLMS adds noise rather than subtracting it from the target signal. The first row in Table 1 presents these statistics. The NLMS cancellation performance points out two significant aspects. Firstly, it outperforms the UWF method in the 10 -- 15\,Hz band approximately 70\% of the time. This is typically due to the stochastic nature of the algorithm, albeit at the cost of performance in the 15 -- 20 \,Hz band. Secondly, there are instances when $r^{\text{NLMS}}_{20, 25}$ exceeds zero (about 8\% of the time) which is undesirable. This typically results from the fixed step-size in the NLMS algorithm. For this reason, several variable step-size NLMS algorithms have been proposed over the last few decades \citep{harris1986variable, kwong1992variable, mathews1993stochastic, aboulnasr1997robust}. However, most of the methods heavily rely on the input signals' statistics and the expected error signal variance. Consequently, our focus is on a specific variant -- a blend of the NLMS and the proportionate NLMS algorithm \citep{duttweiler2000proportionate}. This variant is less dependent on input statistics, offering potential advantages for our application.

\subsection{Improved proportionate NLMS}
The proportionate NLMS (PNLMS) algorithm was developed in the early 2000s for addressing network echo cancellation problems. In particular this method finds wide usage in situations where the echo paths are sparse. Unlike the NLMS algorithm which uses a fixed adaptation step size, the PNLMS algorithm assigns the step size to each of the filter coefficients based on their values in the previous iteration. This can also be visualized as a strategy where larger coefficients receive larger increments at each iteration. It has been proven that the PNLMS algorithm shows faster convergence compared to the NLMS algorithm in problems where the filter coefficients are sparse \citep{duttweiler2000proportionate}. However, in situations where the nature of the filter coefficients are unknown, the PNLMS performs worse compared to the NLMS algorithm \citep{benesty2002improved}. Hence, the improved PNLMS (IPNLMS) which is a mix of NLMS and PNLMS was developed that would perform better than the NLMS algorithm irrespective of the nature of the filter coefficients. The filter coefficients are updated as
\begin{equation}
    \mathbf{h}_{n} = \mathbf{h}_{n-1} + \alpha \frac{\mathbf{X}_{n}^{\dagger}\mathbf{G}_{n-1}\mathcal{E}_{n}}{\mathbf{X}^{\dagger}_{n}\mathbf{G}_{n-1}\mathbf{X}_{n}+\delta_{\text{IPNLMS}}},
    \label{eqn9}
\end{equation}
where $\mathbf{G}_{n-1}$ is a diagonal matrix of size $(NP\times NP)$ at time index $(n-1)$. Each diagonal element $g_l$ corresponding to the $l^{th}$ filter coefficient $h_{l,n-1}$ is expressed as
\begin{equation}
g_{l,n-1} = \frac{1-\beta}{2NP} + (1+\beta)\frac{\mid h_{l,n-1}\mid}{2\|\mathbf{h}_{n-1}\|_1 + \epsilon},
\label{eqn10}
\end{equation}
where $\mid \cdot \mid$ denotes the absolute value, $\|\cdot\|_1$ denotes the $L_1$-norm, and $\epsilon$ is a small positive constant to avoid division by zero. The quantity $\delta_{\text{IPNLMS}}$ in equation (\ref{eqn9}) is computed as
\begin{equation}
    \delta_{\text{IPNLMS}} = \rho\sigma_X^2\frac{(1-\beta)}{2NP},
    \label{eqn11}
\end{equation}
where $\rho$ is a small positive constant and $\sigma_X^2$ is the power of the input signal $\mathbf{X}_{n}$. Examining equations (\ref{eqn9}), (\ref{eqn10}), and (\ref{eqn11}), it becomes evident that the filter coefficient adaptation simplifies to the NLMS algorithm when $\beta = -1$ and the PNLMS algorithm when $\beta = 1$. Similar to the NLMS algorithm, the IPNLMS algorithm is initialized with $\mathbf{h}_{-1} = \mathbf{0}_{1\times NP}$ and $\alpha = 1.0$. The small positive constant $\epsilon$ is set to $10^{-10}$ and $\rho$ is set to 0.01. As previously mentioned, the choice of $\beta$ determines whether the filter adaptation follows the NLMS or PNLMS approach. For most applications, $\beta$ is commonly selected as -0.5 or 0.0, as indicated in \citep{loganathan2010performance}. In our specific application, we opted for $\beta = 0.0$ as it demonstrated better convergence speed and steady-state tracking compared to the NLMS algorithm.

\begin{figure}[ht!]
\begin{center}
    \includegraphics[width=0.48\textwidth]{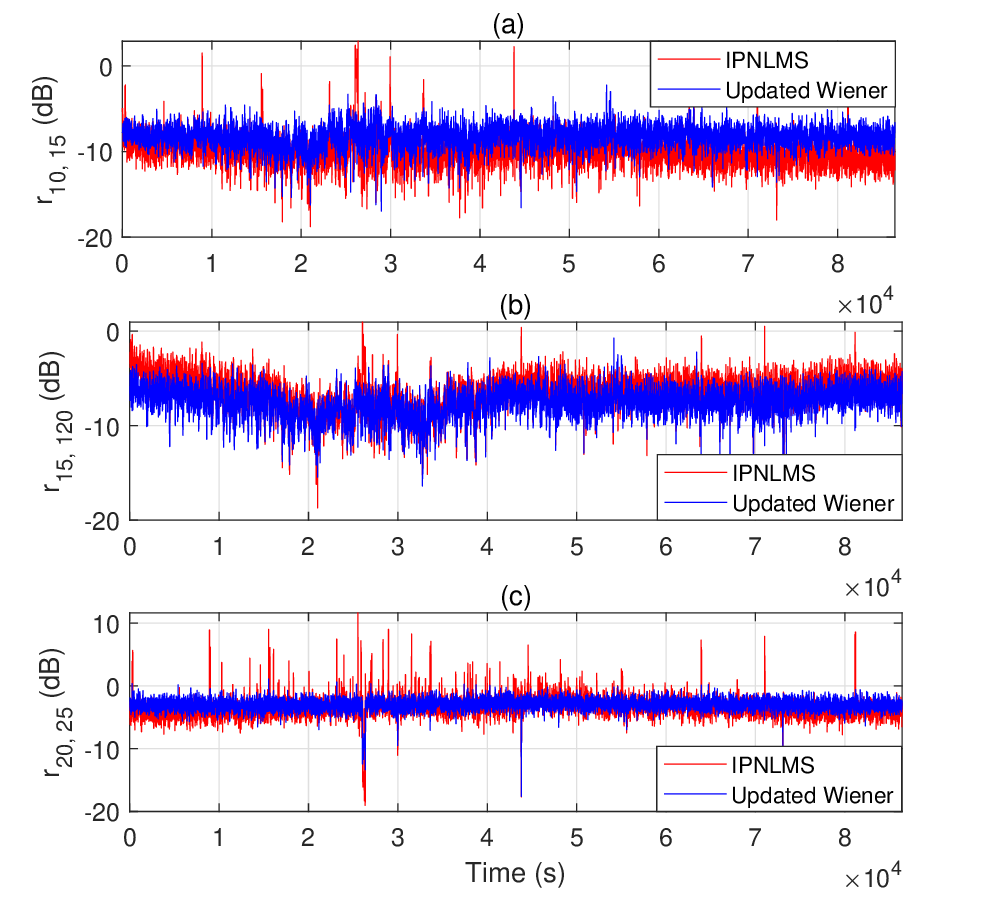}
    \caption{Comparison of noise cancellation performance between the IPNLMS algorithm (red curve) and the UWF method (blue curve) for August 01, 2023 data, across frequency bands: (a) 10 -- 15\,Hz, (b) 15 -- 20\,Hz, and (c) 20 -- 25\,Hz.}
    \label{IPNLMSUWFCompare}
\end{center}
\end{figure}

Similar to the tests performed for the NLMS algorithm, the IPNLMS algorithm was applied to continuous data between August 01 and August 07, 2023. Figures \ref{IPNLMSUWFCompare}(a), (b), and (c) show the comparison of the subtraction performance between the IPNLMS and UWF algorithms for a day of data (August 01, 2023) corresponding to the frequency bands of 10 -- 15\,Hz, 15 -- 20\,Hz, and 20 -- 25\,Hz, respectively. The IPNLMS algorithm outperforms the UWF method in the 10 -- 15\,Hz band for approximately 90\% of the time which is a substantial improvement compared to the NLMS algorithm (about 70\% of the time). Similarly, in the 20 -- 25\,Hz band a better cancellation performance was observed over the UWF method for about 70\% of the time (about 45\% for the NLMS algorithm). Another noteworthy improvement lies in the reduction of time when the algorithm introduces noise to the target instead of subtraction. In the frequency band of 20 -- 25\,Hz, the NLMS algorithm added noise approximately 8\% of the time which is reduced to about 3\% with the implementation of the IPNLMS algorithm. Table \ref{table1} lists several of the aforementioned performance indicators.

\begin{figure}[ht!]
\begin{center}
    \includegraphics[width=0.48\textwidth]{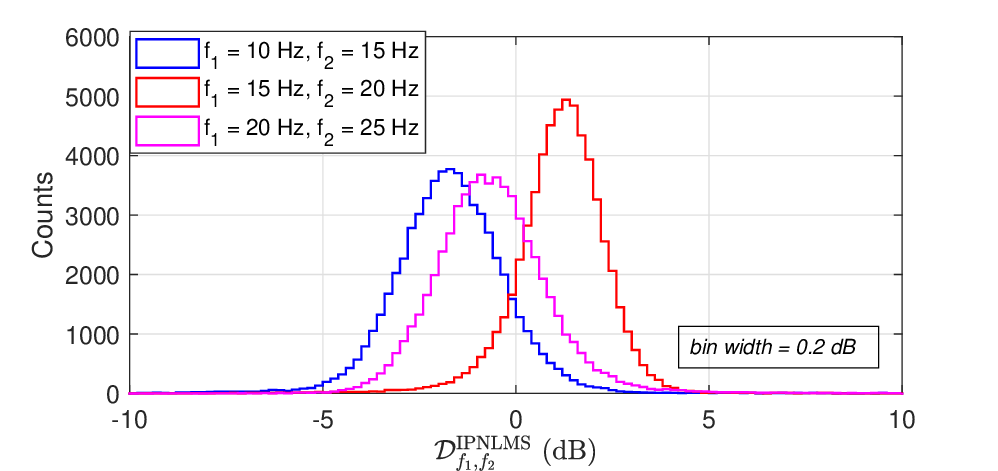}
    \caption{Comparison of the histograms of $\mathcal{D}^{\text{IPNLMS}}_{f_1, f_2}$ corresponding to the frequency bands 10 -- 15\,Hz, 15 -- 20\,Hz, and 20 -- 25\,Hz.} 
    \label{HistIPNLMSUWF}
\end{center}
\end{figure}

In summary, the IPNLMS algorithm exhibits better performance compared to the NLMS algorithm in both the 10 -- 15\,Hz and 20 -- 25\,Hz frequency bands. The histograms of $\mathcal{D}_{f_1, f_2}^{\text{IPNLMS}}$ for these frequency bands are presented in Figure \ref{HistIPNLMSUWF}. However, similar to the NLMS algorithm, the subtraction performance of the IPNLMS algorithm lags behind the UWF method in the 15 -- 20\,Hz frequency band. In order to improve the cancellation performance within the 15 -- 20\,Hz band and minimize the instances of noise addition to the target channel, the next section explores noise cancellation algorithms within the RLS class. Among the various options, including the prewindowed approach \citep{lee1981recursive}, the sliding window method \citep{porat1983normalized}, and the exponentially windowed scheme \citep{cioffi1984fast}, we opt for the latter as it proves effective in tracking changes within a time-varying system.

\begin{table*}[t]
  \centering
  \begin{tabular}{c|c|c|c|c|c|c|c|c|c}
  \hline
  \multicolumn{1}{c|}{\multirow{2}{*}{Methods}} & \multicolumn{9}{c}{Percentage of population}\\
  \cline{2-10}
  \multicolumn{1}{c|}{} & $\mathcal{D}_{10, 15}\leq0$ & $\mathcal{D}_{10, 15}\leq2$ & $r_{10, 15}>0$ & $\mathcal{D}_{15, 20}\leq0$ & $\mathcal{D}_{15, 20}\leq2$ & $r_{15, 20}>0$ & $\mathcal{D}_{20, 25}\leq0$ & $\mathcal{D}_{20, 25}\leq2$ & $r_{20, 25}>0$
  \\
  \hline
  \multicolumn{1}{c|}{NLMS} & 71.3 & 95.58 & 0.59 & 8.09 & 74.31 & 0.83 & 45.26 & 86.42 & 7.93
  \\
  \hline
  \multicolumn{1}{c|}{IPNLMS} & 89.67 & 98.91 & 0.24 & 12.86 & 78.98 & 0.20 & 69.83 & 95.92 & 3.00
  \\
  \hline
  \multicolumn{1}{c|}{FTF-RLS} & 54.56 & 99.55 & 0.36 & 41.38 & 99.56 & 0.37 & 27.76 & 99.26 & 0.68
  \\
  \hline
  \end{tabular}
  \caption{Performance statistics of the NLMS, IPNLMS, and FTF-RLS algorithms concerning the parameters $\mathcal{D}_{f_1,f_2}$ and $r_{f_1,f_2}$ for the frequency bands of 10 -- 15\,Hz, 15 -- 20\,Hz, and 20 -- 25\,Hz. The statistics were derived from 60,480 observations, covering the period from August 01, 2023, to August 07, 2023.}
  \label{table1}
\end{table*}

\section{Recursive Least Square Algorithm}
\label{sec:RLS}
In the Recursive Least Square (RLS) algorithm at time index $n$, the filter coefficients $\mathbf{h}_{n}$ are obtained by solving
\begin{equation}
    \min_{\mathbf{h}_{n}}\sum_{t=0}^{n}\lambda^{n-t}e_n e_{n}^{\dagger},
    \label{eqn12}
\end{equation}
where $\lambda$ is called the forgetting factor and $e_n$ follows from equation (\ref{eqn1}). The value of $\lambda$ is set to a value very close to 1.0. This is in contrast to the prewindowed scheme where lambda is precisely set to 1.0. Following equation (\ref{eqn12}), this implies that with growing number of iterations the impact of the errors from the past in determining the current values of the filter coefficients gradually diminish. For any time index $n>NP$, the filter coefficients can be obtained by using the following recursions \citep{slock1989fast},
\begin{align}
    \tilde{\mathbf{C}}_{n} &= \mathbf{X}_n^{\dagger}\lambda^{-1}\mathbf{R}_{n-1}^{-1} \label{eqn13}\\
    \gamma_n^{-1} &= 1+\tilde{\mathbf{C}}_{n}\mathbf{X}_{n}\label{eqn14}\\
    e_{n}^p &= y_n - \mathbf{h}_{n-1}\mathbf{X}_{n} \label{eqn15}\\
    \mathbf{h}_{n} &= \mathbf{h}_{n-1} + \gamma_n e_n^p\tilde{\mathbf{C}}_{n} \label{eqn16}\\
    \mathbf{R}_{n}^{-1} &= \lambda^{-1}\mathbf{R}_{n-1}^{-1} - \tilde{\mathbf{C}}_{n}^{\dagger}\gamma_n\tilde{\mathbf{C}}_{n}. \label{eqn17}
\end{align}
The filter coefficients are updated using equation (\ref{eqn16}), where the update involves the normalized Kalman gain $\tilde{\mathbf{C}}_{n}$ and the apriori error estimate $e_{n}^p$. The scaling factor $\gamma_{n}$ in equation (\ref{eqn14}) can be expressed equivalently as $\tilde{\mathbf{C}}_{n} = \gamma_{n}^{-1}\mathbf{C}_{n}$, where $\mathbf{C}_{n}$ represents the unnormalized Kalman gain. The Kalman gain $\tilde{\mathbf{C}}_{n}$ is estimated using equation (\ref{eqn13}) where
\begin{equation}
    \mathbf{R}_{n} = \sum_{t=0}^n\lambda^{n-t}\mathbf{X}_{n}\mathbf{X}^{\dagger}_{n}
\end{equation}
represents the exponentially weighted input data covariance matrix. Recursive estimation of the inverse of matrix $\mathbf{R}_{n}$ in equation (\ref{eqn17}) follows from the `Sherman–Morrison–Woodbury matrix inversion lemma' \citep{woodbury1950inverting}. Similar to the LMS implementations, the filter coefficients at the start of the algorithm are initialized as $\mathbf{h}_{-1} = \mathbf{0}_{1\times NP}$, and the matrix $\mathbf{R}_{-1}^{-1}$ is set to $\{\mu \mathbf{\Lambda\}}^{-1}$, where $\mu$ is a positive scalar weighting factor and $\mathbf{\Lambda} = \text{diag}\{ \lambda^{NP},\lambda^{NP-1},\cdots, \lambda \}$. Following equation (4.17) in \citep{cioffi1984fast}, a reasonable choice for the value of $\mu$ is $\frac{L\sigma_X^2}{NP}$ where $L$ is the number of data points over which the input signal power $\sigma_X^2$ is estimated.

The matrix $\mathbf{R}_{n}$ and the Kalman gain vector $\tilde{\mathbf{C}}_{n}$ in equations (\ref{eqn13}) -- (\ref{eqn17}) have dimensions $(NP\times NP)$ and $(1\times NP)$, respectively. Consequently, the computational complexity per time update of the RLS algorithm amounts to about $3N^2P^2 + 5NP$. In our specific implementation with $N = 101$ and $P = 24$, this level of computational complexity for low-latency implementation is impractical. Moreover, future implementation of the NN cancellation system at Virgo is expected to use more than 100 witness channels. Hence, in order to address the computational load, we aim to tackle the RLS problem using faster algorithms without compromising on the performance.

Two main classes of fast algorithms address the RLS problem: the fast lattice and the fast transversal filter (FTF) algorithms. Early works on the development of the lattice algorithms can be found in \citep{lee1981recursive, friedlander1982lattice}. Although these algorithms have a computational complexity of about $O(NP^2)$, they significantly outpace the standard RLS algorithm. Subsequent improvements in speed over the lattice algorithms were achieved with the development of the FTF versions. Some examples of the early FTF versions include the fast Kalman \citep{ljung1978fast, falconer1978application} and the FAEST algorithms \citep{carayannis1983fast}. While these algorithms also exhibit a $O(NP^2)$ complexity, the coefficient of $NP^2$ is considerably smaller compared to the lattice versions. Consequently, we chose to proceed with the FTF version of the fast algorithms.

The time update of the filter coefficients in the FTF-RLS algorithm follows the same as in equation (\ref{eqn16}). However, the update of the Kalman gain $\tilde{\mathbf{C}}_{n}$ does not follow the standard implementation. Instead a different time-updating scheme is used. These are commonly referred to as the order update and down date procedures. The derivation of the FTF-RLS algorithm is more tedious and complex as compared to the standard RLS implementation. The FTF-RLS algorithm used in our implementation can be found in \citep{slock1991numerically}. One of the problems identified within a few years of development of the FTF algorithms was related to their numerical stability. These algorithms were found to be exponentially unstable \citep{ljung1985error,ardalan1987fixed}, implying that the filter coefficients could diverge after a certain number of iterations. While operating with infinite precision would prevent this instability, the algorithm used in this article \citep{slock1991numerically} tackles numerical instability by leveraging redundancies. During the Kalman gain update, identical parameters are computed through different formulations and convex combinations of these estimates are fed back at various stages in the algorithm. Detailed information regarding the optimal feedback coefficients is provided in Section VI of \citep{slock1991numerically}. In our implementation of the algorithm, we have used the same values of the coefficients as proposed in \citep{slock1991numerically}. Considering the intricacies involved in implementing this algorithm and to ensure brevity for readers, as well as accommodate potential future upgrades to the technique, the computer programs are available at \citep{koleyAdapNNC}. For easier comprehension, we have used the same variable notations as outlined in Table II of \citep{slock1991numerically}.

An important parameter that we have not discussed so far and which impacts the stability of the FTF-RLS algorithm is the choice of the forgetting factor $\lambda$. Based on studies in \citep{slock1989fast, benallal1989improvement}, the condition $1 > \lambda > 1 - \frac{1}{mNP}$, where $m>2$ must be satisfied.  Opting for a value of $\lambda$ very close to 1.0 ensures numerical stability. However, this occurs at the cost of slower convergence given that the time constant of the FTF-RLS algorithm can be expressed as $\frac{1}{1-\lambda}$ \citep{eleftheriou1986tracking}. Hence, in order to obtain a right balance between stability and speed of convergence we chose $\lambda = 1-\frac{1}{3NP}$ where the time constant $3NP$ is expressed as number of samples.

\begin{figure}[ht!]
\begin{center}
    \includegraphics[width=0.48\textwidth]{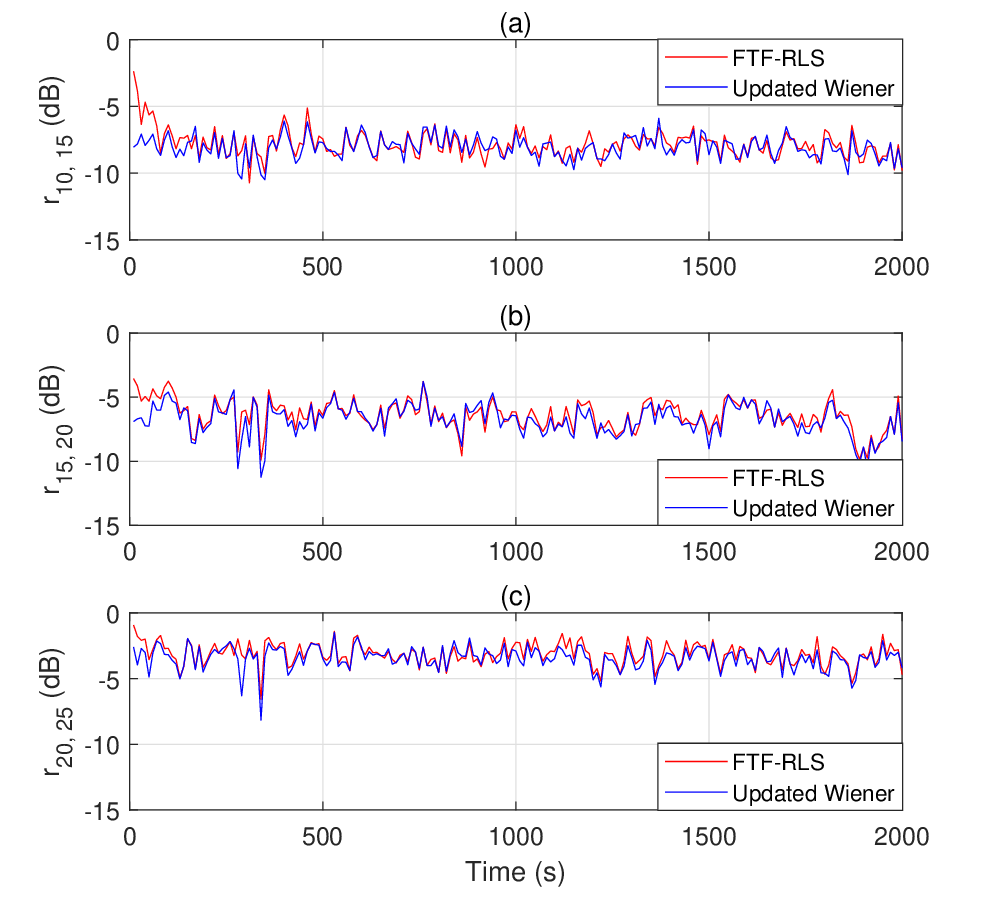}
    \caption{The red curve shows the convergence of the FTF-RLS algorithm to the UWF's cancellation performance (blue curve) within the first 80 -- 100\,s after start corresponding to the frequency bands (a) 10 -- 15\,Hz. (b) 15 -- 20\,Hz, and (c) 20 -- 25\,Hz.}
    \label{FTFRLSConv}
\end{center}
\end{figure}

Using the values of $\mu$ and $\lambda$ as stated earlier and following the same pre-processing steps implemented in the UWF and LMS schemes, we applied the FTF-RLS algorithm to continuous data measured between August 01 - 07, 2023. Figures \ref{FTFRLSConv}(a), (b), and (c) present a performance comparison between the FTF-RLS and the UWF method for the initial 2000\,s following the start of the FTF-RLS algorithm. Convergence of the subtraction performance to the UWF method occurs in about 100\,s after the start. Ideally for uncorrelated inputs, convergence should occur in $3NP$ samples which corresponds to about 72\,s after the start ($P = 24, N = 101$, and sampling rate of 100\,Hz). However, in our application the inputs are correlated. Hence, the convergence is delayed. Most of the analysis on convergence and steady state performance of the FTF algorithms have been performed for white Gaussian input sequences \citep{eleftheriou1986tracking}. As we note, for applications with correlated inputs, the observed and the theoretical values of convergence time differ. Nevertheless, the algorithm was numerically stable and was found to run seamlessly for days.

\begin{figure}[ht!]
\begin{center}
    \includegraphics[width=0.48\textwidth]{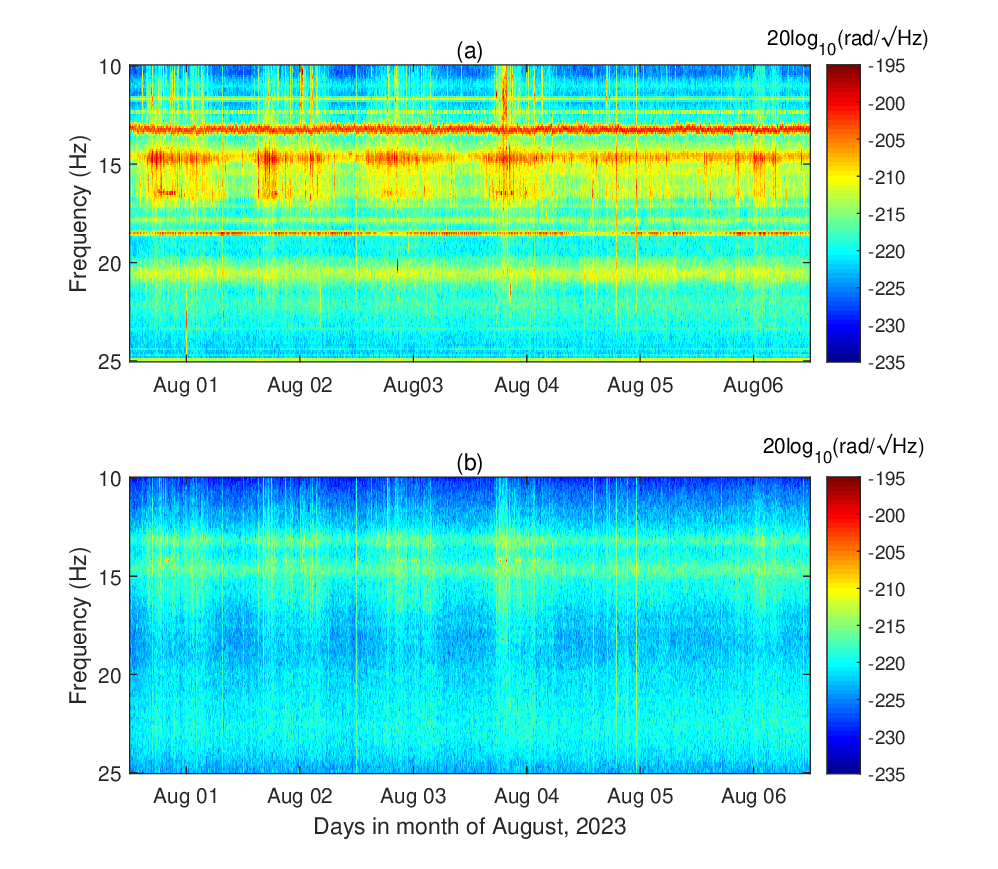}
    \caption{Spectrogram of the (a) target data and (b) FTF-RLS cleaned data for the period from August 01 to August 06, 2023. Average PSDs are computed every 100\,s using window lengths of 10\,s, with an overlap of 5\,s between successive time windows.}
    \label{FTFRLSSpect}
\end{center}
\end{figure}

Figures \ref{FTFRLSSpect}(a) and (b) display the spectrograms illustrating the target data and the FTF-RLS cleaned data for the period spanning August 01 to August 06, 2023. The temporal resolution of the spectrogram is 100\,s. Power spectral densities are computed at 100\,s intervals, employing a Hann window with a length of 10\,s and an overlap of 5\,s between successive windows. A strong noise subtraction of about  20 -- 25\,dB is observed for sharp spectral noise peaks at frequencies such as 11.6\,Hz, 12.3\,Hz, 13.4\,Hz, and 18.5\,Hz. In the case of broadband noise, a weaker subtraction of about 10\,dB is observed. This performance aligns with the observed correlations between the witness and the target signals. Strong correlations exceeding 0.8 were observed for sharp spectral peaks, while broadband noise had correlations between 0.2 and 0.4 (Figure \ref{NEBTiltCC}).

\begin{figure}[ht!]
\begin{center}
    \includegraphics[width=0.48\textwidth]{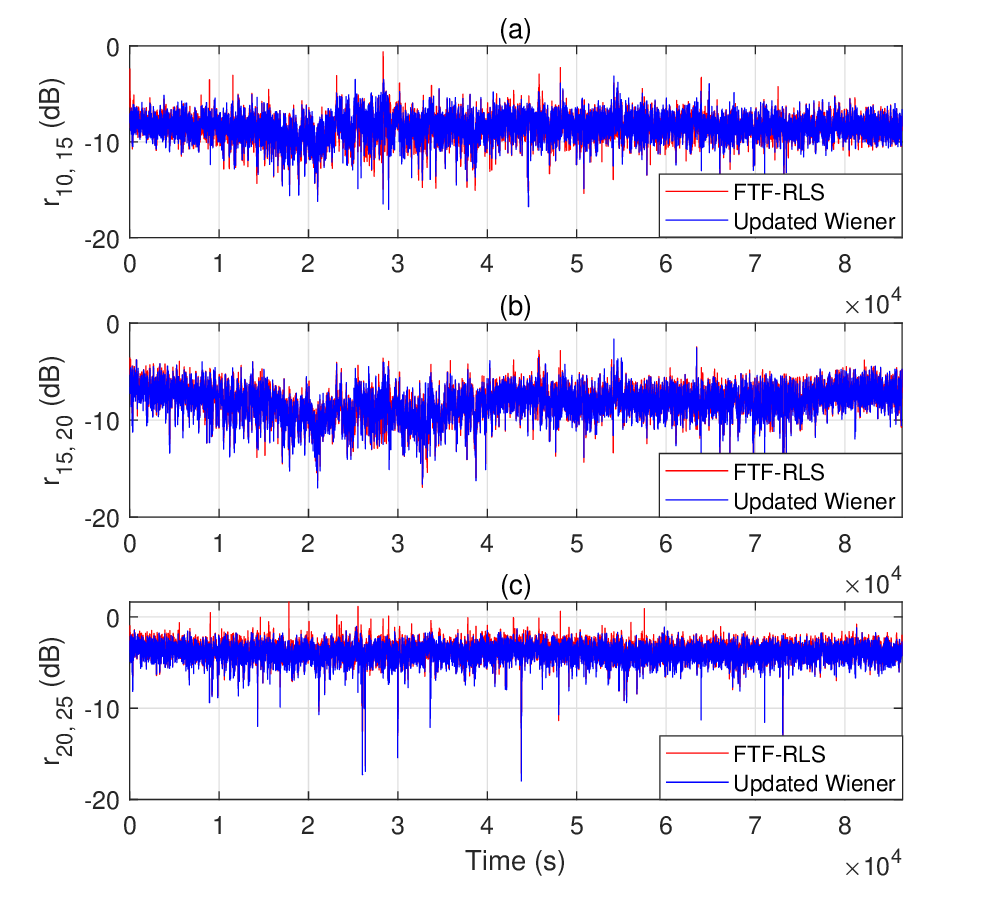}
    \caption{Comparison of noise cancellation performance between the FTF-RLS algorithm (red curve) and the UWF method (blue curve) for August 01, 2023 data, across frequency bands: (a) 10 -- 15\,Hz, (b) 15 -- 20\,Hz, and (c) 20 -–25\,Hz.}
    \label{FTFRLSUWFComp}
\end{center}
\end{figure}

\begin{figure}[ht!]
\begin{center}
    \includegraphics[width=0.48\textwidth]{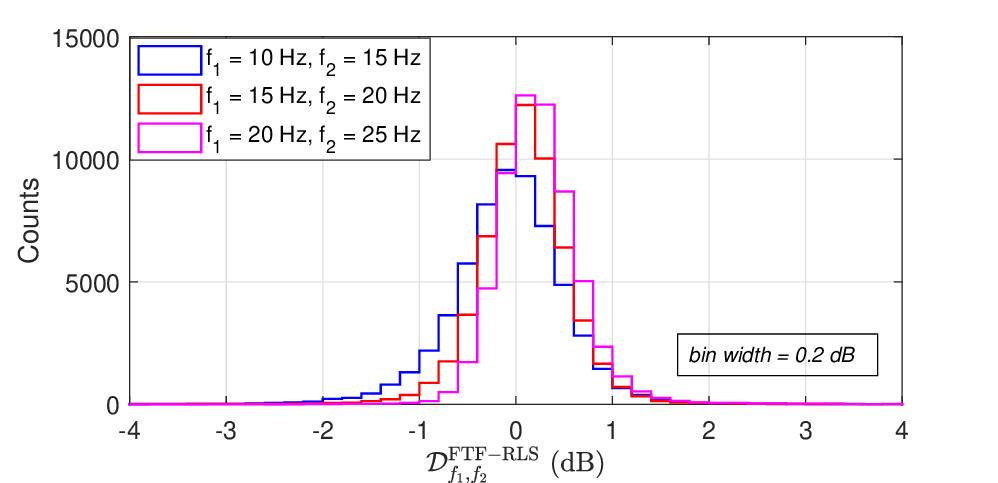}
    \caption{Comparison of the histograms of $\mathcal{D}^{\text{FTF-RLS}}_{f_1, f_2}$ corresponding to the frequency bands 10 -- 15\,Hz, 15 -- 20\,Hz, and 20 -- 25\,Hz.} 
    \label{FTFRLSHist}
\end{center}
\end{figure}

In line with the noise-cancellation tests performed for the UWF and LMS algorithms, we calculated $r_{f_1,f_2}^{\text{FTF-RLS}}$ for the frequency bands of 10 -- 15\,Hz, 15 -- 20\,Hz, and 20 -- 25\,Hz. Figures \ref{FTFRLSUWFComp}(a), (b), and (c) present the comparison of the noise cancellation efficiency between the FTF-RLS and UWF across these three frequency bands. The performance of the FTF-RLS algorithm matches the UWF method across all three frequency bands. This is in contrast to the LMS method which exhibited bias in the 15 -- 20\,Hz range.  Over a continuous seven-day run, the FTF-RLS algorithm achieves a performance within 2\,dB of the cancellation achieved by the UWF method for more than 99\% of the time. Another improvement is evident in the reduction of the percentage of time the cancellation algorithm introduces noise instead of subtraction in the 20 -- 25\,Hz band. This is reduced to below one percent of the time. The histograms of $\mathcal{D}_{f_1,f_2}^{\text{FTF-RLS}}$ corresponding to the frequency bands 10 -- 15\,Hz, 15 -- 20\,Hz, and 20 -- 25\,Hz are shown in Figure \ref{FTFRLSHist}. Similar to the LMS tests, the total number of observations are 60,480 corresponding to the seven day analysis window between August 01 and August 07, 2023. Each of the three distributions are centered around zero, which is an improvement over the LMS performance. All the three distributions also have a much smaller standard deviation compared to that observed in the LMS implementations. The third column in Table \ref{table1} details several of these performance statistics.

Based on the performance statistics presented in Table \ref{table1} of the three different adaptive noise-cancellation methods, we find that the FTF-RLS method performs the best. Moreover, the time complexity is also suitable for low-latency applications. However, there are certain avenues of improvement that have not been addressed so far, and we present these possibilities in the next section.

\section{Future developments}
\label{sec:FD}
All of the adaptive noise-cancellation methods that have been implemented in this article perform the best when the inputs are uncorrelated. In a Virgo-like seismic environment where all the input channels are located within tens of meters of each other, strong correlations between the inputs are observed at lower frequencies while correlations are weaker at higher frequencies (the turning point between these two regimes is around 15\,Hz \cite{harms2019terrestrial}). This necessitates the exploration of decorrelation techniques for reducing the correlation between nearby input channels. A simple nonlinear method is that of a half-wave rectifier \citep{benesty1998better} so that the nonlinearly transformed signal becomes
\begin{equation}
    x'(n) = x(n) + \kappa\frac{x(n) + \mid x(n)\mid}{2},
    \label{eqn19}
\end{equation}
where $\kappa$ is a parameter used to control the nonlinearity. An improved version of the decorrelation technique that makes use of alternating positive and negative half-wave rectifiers have been discussed in Chapter 1 in \citep{benesty2004adaptive}. A performance analysis of several other methods like the `Hard Limiter', `Square-Sign', and `Square-Law' for introducing nonlinearities between the input signals can be found in \citep{morgan2001investigation}. However, utmost care must be taken while introducing these nonlinearities, which might as well degrade the cancellation performance. As an alternative to the generic decorrelation technqiues, one might explore solutions tailored to the NNC case. However, while a frequency-dependent decorrelation is easy to design in frequency domain, it is challenging to solve the problem in the context of causal time-domain filters as needed for NNC.

An assumption under which the FTF-RLS algorithm was developed was that the input signals are `persistently exciting'. This condition again points to the problem associated with the non-whiteness of the input signals. The algorithm used in this article is based on a soft-constrained rescue mechanism, that handles the situation of eventual instability. The instability can be attributed to the condition number of some of the matrices becoming very large. In the current re-initialization method, these matrices are reset to the values as if the algorithm were started for the first time. Although, the other variables like the filter coefficients are retained, the re-initialization leads to suboptimal performance of the algorithm for a few seconds. In our seven day continuous run, this was encountered four times, and was not a huge problem. However, some researchers have made use of a mix of NLMS and FTF-RLS for handling such situations \citep{eleftheriou1984restart}. They switch to the NLMS algorithm for filter updates at the time when the FTF-RLS encounters instability. After a few hundred seconds when the FTF-RLS has stabilized, the method switches back to FTF-RLS instead of the NLMS algorithm. We did not implement such a mixed scheme, but it is something that could be explored.

Finally a detailed study of the impact of these adaptive noise subtraction schemes on gravitational-wave searches needs to be performed. Even if an adaptive NNC reduces noise on average, its effect on the transient background might be different. In addition, according to equation (\ref{eqn6}) and similarly for all adaptive algorithms, the filter itself is susceptible to transients in the data (in the target as well as witness channels), which is most relevant to filters based on the stochastic gradient descent. It will be important to carefully characterize adaptive Wiener filters in terms of their effect on the transient background of the target channel.

\section{Conclusion}
\label{sec:conclusion}
In this paper, we analyzed algorithms for adaptive Wiener filtering. We found that they all outperform the static Wiener filter. The reason for the advantage of adaptive filters is that the properties of the seismic field at the Virgo site change with time. Most importantly, the day-night cycle must be tracked by the filter for improved performance. All adaptive algorithms perform similarly even though the RLS algorithm had the most consistent performance across the entire NN band from 10\,Hz -- 25\,Hz.

We discussed fundamental performance limitations of noise cancellation with Wiener filters and derived a lower limit on the residuals due to filter bias from statistical errors in the correlation estimates. This lower limit becomes more stringent with increasing number of filter coefficients, which puts in question NNC strategies relying on an increasing number of sensors and increasingly complex noise-cancellation filters. Mathematics rewards economical filter designs.

At this point, the adaptive algorithms are understood well enough to implement them in NNC systems. However, their effect on the detector data must be investigated. The filters are designed to provide a noise reduction on average, but their impact on the transient noise background is unknown. The adaptive filter itself can be disturbed by transients in the data. 

Finally, there are several efforts to introduce machine-learning algorithms for noise cancellation. While these methods obey the same limits on noise residuals and in most cases increase the complexity of the training compared to adaptive Wiener filters, there might be interesting applications when it comes to clever adaptation to more complex time-variations of the seismic field. For example, a seismic field might have different states that repeat and one could imagine to switch between different filters adapted to the different states. This can in principle be done with Wiener filters as well, but it might be possible to realize it as a fully automatic process with machine learning.

\section{Acknowledgements}
The authors gratefully acknowledge the Italian Istituto Nazionale di Fisica Nucleare (INFN), the French Centre National de la Recherche Scientifique (CNRS) and the Netherlands Organization for Scientific Research (NWO), for the construction and operation of the Virgo detector and the creation and support of the EGO consortium. The authors also gratefully acknowledge research support from these agencies as well as by the Spanish Agencia Estatal de Investigaci\'on, the Consellera d’Innovaci\'o, Universitats, Ci\`encia i Societat Digital de la Generalitat Valenciana and the CERCA Programme Generalitat de Catalunya, Spain, the National Science Centre of Poland and the European Union—European Regional Development Fund; Foundation for Polish Science (FNP), the Hungarian Scientific Research Fund (OTKA), the French Lyon Institute of Origins (LIO), the Belgian Fonds de la Recherche Scientifique (FRS-FNRS), Actions de Recherche Concert\'ees (ARC) and Fonds Wetenschappelijk Onderzoek—Vlaanderen (FWO), Belgium, the European Commission. The authors gratefully acknowledge the support of the NSF, STFC, INFN, CNRS and Nikhef for provision of computational resources.

Soumen Koley acknowledges the support through a collaboration agreement between Gran Sasso Science Institute and Nikhef and from the European Gravitational Observatory through a collaboration convention on Advanced Virgo +. The authors also gratefully acknowledge the support of the Italian Ministry of Education, University and Research within the PRIN 2017 Research Program Framework, n. 2017SYRTCN.

\bibliographystyle{apsrev}
\bibliography{references}

\end{document}